\documentclass[aps,prd,superscriptaddress,eqsecnum,showpacs]{revtex4}
\pdfoutput=1
\usepackage{graphicx}
\usepackage{color}
\usepackage{amssymb}
\usepackage{epstopdf}
\usepackage[colorlinks,urlcolor=blue,citecolor=blue]{hyperref}
\def\hhref#1{\href{http://arxiv.org/abs/#1}{arXiv:#1}} 

\begin{document}

\title{Full Time-Dependent Hartree-Fock Solution of Large N Gross-Neveu  models}
\author{Gerald V. Dunne}
\affiliation{ARC Centre of Excellence in Particle Physics at the Terascale and CSSM,
School of Chemistry and Physics, University of Adelaide, Adelaide SA 5005, Australia}
\affiliation{Physics Department, University of Connecticut, Storrs CT 06269, USA}
\author{Michael Thies}
\affiliation{Institut f\"ur  Theoretische Physik, Universit\"at Erlangen-N\"urnberg, D-91058,
Erlangen, Germany }

\begin{abstract}

We find the general solution to the time-dependent Hartree-Fock problem for scattering solutions of the Gross-Neveu models, with 
both discrete (GN$_2$) and continuous
(NJL$_2$) chiral symmetry. We find new multi-breather solutions both for the GN$_2$ model, generalizing the Dashen-Hasslacher-Neveu 
breather solution, and also new
twisted breathers for the NJL$_2$ model. These solutions satisfy the full TDHF consistency conditions, and only in the special
cases of GN$_2$ kink scattering do these conditions 
reduce to the integrable Sinh-Gordon equation. We also show that all baryons and breathers are composed of constituent twisted
kinks of the NJL$_2$ model. Our solution depends crucially on a general class of transparent, time dependent Dirac potentials 
found recently by algebraic methods.

\end{abstract}

                                          
\maketitle

\section{Introduction}

The  Gross-Neveu (${\rm GN}_2$) and Nambu-Jona Lasinio (${\rm NJL}_2$) models in 1+1 dimensional quantum field theory
describe $N$ species of massless, self-interacting Dirac fermions with Lagrangians \cite{Gross:1974jv}: 
\begin{eqnarray}
{\cal L}_{\rm GN} &=& \sum_{k=1}^N \bar{\psi}_k i\partial \!\!\!/ \psi_k + \frac{g^2}{2} \left( \sum_{k=1}^N \bar{\psi}_k \psi_k \right)^2
\label{gn}\\
{\cal L}_{\rm NJL} &=&  \sum_{k=1}^N \bar{\psi}_k i\partial \!\!\!/ \psi_k + \frac{g^2}{2} \left[\left( \sum_{k=1}^N \bar{\psi}_k \psi_k \right)^2 +
\left( \sum_{k=1}^N \bar{\psi}_k i \gamma_5\psi_k \right)^2\right].
\label{njl}
\end{eqnarray}
These models  serve as soluble paradigms for symmetry breaking phenomena in both strong interaction particle physics and in 
condensed matter physics \cite{Nambu:1961tp,Thies:2006ti}. 
We consider these models in the 't~Hooft limit, $N \to \infty$, with $Ng^2=$ constant, where semiclassical methods become exact.
Classically, the ${\rm GN}_2$ model has a discrete chiral symmetry, while the ${\rm NJL}_2$ model has a continuous chiral symmetry. At  
finite temperature and density these models exhibit a rich structure of phases with inhomogeneous  crystalline condensates in the
large $N$ limit, these phases being directly associated with chiral symmetry breaking  \cite{Basar:2009fg}. 
Such self-interacting fermion models also have numerous applications to a wide variety of phenomena in particle, condensed matter and atomic physics 
\cite{Fulde:1964zz,peierls,horovitz,braz,Okuno:1983,machida,machida2,machida3,rajagopal,casalbuoni,Heeger:1988zz,machida-amo,zwierlein,pitaevskii,Adams:2012th,Kanamoto:2008zz,Herzog:2007ij}.

 In the 't~Hooft limit, 
$N \to \infty$,  $Ng^2=$ constant, we use semi-classical techniques pioneered in this context by Dashen, Hasslacher and Neveu (DHN)
\cite{Dashen:1974ci}. This can either be understood 
in functional language as a gap equation, or as a Hartree-Fock problem in which one solves the Dirac equation subject to constraints
on the scalar and pseudoscalar condensates. Here we
use the time-dependent Hartree-Fock (TDHF) formalism, which involves solving the following constrained Dirac equations:
\begin{eqnarray}
{\rm GN}_2 &:&\quad (i \partial \!\!\!/ - S(x, t))\psi_{\alpha} = 0\quad , \quad S=-g^2 \sum_{\beta}^{\rm occ} \bar{\psi}_{\beta} \psi_{\beta}
\label{gn-tdhf}
\\
{\rm NJL}_2 &:&\quad (i \partial \!\!\!/ - S(x, t) -i \gamma_5 P(x, t))\psi_{\alpha} = 0\quad , \quad S=-g^2 \sum_{\beta}^{\rm occ} \bar{\psi}_{\beta} 
\psi_{\beta} \quad, \quad P=-g^2 \sum_{\beta}^{\rm occ} \bar{\psi}_{\beta}i \gamma_5 \psi_{\beta}
\label{njl-tdhf}
\end{eqnarray}
For NJL$_2$ it is convenient to combine the scalar and pseudo scalar condensates into a single complex condensate 
\begin{eqnarray}
\Delta=S-i P.
\label{eq:delta}
\end{eqnarray}
All {\it static} solutions to these HF problems have been found and used to solve analytically the equilibrium thermodynamic 
phase diagrams of these models in the large $N$ limit, at 
finite temperature and nonzero baryon density  \cite{Thies:2006ti,Basar:2009fg}. These static solutions reveal a deep connection to 
integrable models, in particular the mKdV system for 
the GN$_2$ system, and AKNS for the NJL$_2$ system \cite{Basar:2009fg,Correa:2009xa}. 
In this paper we present a significant extension of these results, by finding the full set of time-dependent solutions to the TDHF equations in 
(\ref{gn-tdhf}) and (\ref{njl-tdhf}) \cite{Dunne:2013xta}. We solve these problems in generality, describing the time-dependent scattering
of non-trivial topological objects such as kinks, 
baryons and breathers. Some special cases have been solved previously, but here we present 
several entirely new classes of solutions to the TDHF problem.
Surprisingly, we have found that the most efficient strategy is to solve the (apparently more complicated) NJL$_2$ model first, and then
obtain GN$_2$ solutions by imposing further constraints on these solutions. For example, we show that the GN$_2$ baryons found by 
Dashen, Hasslacher and Neveu \cite{Dashen:1974ci} can be thought of as bound objects of  twisted NJL$_2$ kinks, and furthermore that 
the scattering of the GN$_2$ baryons can be deduced from the scattering of twisted kinks, a problem whose solution we present here. 
Breathers are somewhat more involved, but again we give a complete and constructive derivation of all multi-breather solutions, also in terms 
of constituent twisted kinks. This 
includes new breather and multi-breather solutions in NJL$_2$, as  well as new multi-breather solutions in the GN$_2$ model.

We stress that while it is well known that the classical equations of motion for the GN$_2$ and NJL$_2$ models  are closely related to
integrable models \cite{Zakharov:1973pp, Pohlmeyer:1975nb,Neveu:1977cr}, this fact is only directly useful for the solution of the 
time-dependent Hartree-Fock problem for the simplest case of kink scattering in the GN$_2$ model, where the problem reduces to 
solving the integrable nonlinear Sinh-Gordon equation \cite{Klotzek:2010gp,Fitzner:2010nv,Jevicki:2009uz}. The more general 
self-consistent TDHF solutions that we find here {\it do not} satisfy the Sinh-Gordon equation, or any simple general bosonic nonlinear 
equation. Instead we shall make use of the transparent, time dependent Dirac potentials derived recently by solving a finite
algebraic problem \cite{Dunne:2013tr}.
We also emphasize that these more general solutions require a self-consistency condition relating the filling fraction of valence fermion
states to the parameters of the condensate solution, 
as for the static GN$_2$ baryon \cite{Dashen:1974ci}, the static twisted kink \cite{Shei:1976mn}, and the GN$_2$ breather
\cite{Dashen:1974ci}. For our time-dependent solutions, this 
important fact means that during scattering processes there is non-trivial back-reaction between fermions and their associated 
condensates and densities 
\cite{Dunne:2011wu}. Kink scattering in the GN$_2$  model,  described by Sinh-Gordon solitons 
\cite{Klotzek:2010gp,Fitzner:2010nv,Jevicki:2009uz}, is much 
simpler because there is no fermion filling-fraction self-consistency condition, nor back-reaction.

\subsection{Basic Building Blocks} 

The  known Hartree-Fock solutions are characterized by several basic building blocks: kinks, baryons, and breathers. We briefly review 
these solutions below. In fact we  show in this
paper that the general solutions are all built out of one basic unit, the twisted kink. To simplify the notation, we henceforth 
set $m=1$, measuring dimensional quantities in terms of the dynamically generated fermion mass $m$.

\begin{enumerate}
\item
{\bf Real CCGZ kink for GN$_2$:} 
The most familiar HF solution for the GN$_2$ model  is the static Coleman-Callan-Gross-Zee (CCGZ) kink \cite{Dashen:1974ci}. Since
we can restrict ourselves to potentials which go to 1 for $x\to - \infty$ without loss of generality, we quote the ``antikink":
\begin{eqnarray}
&&\text{condensate:} \quad S(x)= - \tanh x=\frac{1-e^{2x}}{1+e^{2x}} \label{eq:ccgz}\\
&&\text{fermion filling-fraction consistency condition:} \quad {\rm none}
\nonumber 
\end{eqnarray}
We have expressed the usual tanh form as a ratio of polynomials of exponentials, as this is the basic form of the more general solutions. 
This static kink can be boosted with some velocity  to produce a simple time-dependent solution.

\item
{\bf Complex Twisted Kink for NJL$_2$:} 
The corresponding kink-like solution for the NJL$_2$ model, Shei's twisted kink \cite{Shei:1976mn},  can be expressed in terms of the 
complex condensate $\Delta$ defined in (\ref{eq:delta}):
\begin{eqnarray}
&&\text{condensate:} \quad \Delta(x)= \frac{1+ e^{-2i\theta} e^{2 x \sin \theta}}{1+e^{2x \sin \theta}} \label{eq:shei}
\\
&&\text{fermion filling-fraction consistency condition:} \quad \nu=\frac{\theta}{\pi}
\nonumber
\end{eqnarray}
For $\theta>0$ this kink rotates through an angle $-2\theta$ in the chiral $(S, P)$ plane as $x$ goes from $-\infty$ to $+\infty$. Notice 
that both the 
magnitude, $|\Delta(x)|$, and the phase, $\arg \Delta(x)$, vary with $x$.
When $\theta=\pi/2$, the twisted kink
becomes real, and reduces to the GN$_2$ kink in (\ref{eq:ccgz}). As in  (\ref{eq:ccgz}), the solution can be expressed 
as a rational function of simple exponentials.  This twisted kink solution reveals a new level of complexity, as the self-consistency of 
the HF solution requires a relation between the chiral angle parameter $\theta$ and the fermion filling fraction of  the valence bound 
state \cite{Shei:1976mn}. This fact is responsible for more intricate scattering dynamics of twisted kinks, as  there is a back-reaction from 
the bound fermions during scattering processes, a phenomenon that does not occur for scattering of CCGZ kinks in the GN$_2$ model. 
This is discussed in detail below. Note that the single twisted kink in (\ref{eq:shei}) can also be boosted with some velocity to produce a simple 
time-dependent solution.

\item
{\bf Real DHN Baryon for GN$_2$:}
DHN found a self-consistent static baryon solution for the GN$_2$ model that looks like a bound kink and anti-kink, at locations 
$x=\pm c_0/y$ \cite{Dashen:1974ci}:
\begin{eqnarray}
&&\text{condensate:} \quad S(x)= 1 + y \left[ \tanh(yx-c_0)-\tanh (yx+c_0) \right] \label{eq:dhn-baryon} \\
&&\hskip 3cm =  \frac{1+\frac{ 2\cos 2\theta }{ \cos \theta }e^{2yx}+e^{4yx}}{1+\frac{2}{\cos\theta }e^{2 y x}+e^{4 yx}} \qquad, \qquad 
y = \sin \theta, \quad c_0 = \frac{1}{2} {\rm artanh}\, y \nonumber \\
&&\text{fermion filling-fraction consistency condition:} \quad \nu=\frac{2 \theta}{\pi}
\nonumber
\end{eqnarray}
As $y\to 1$, one or other of the  kink or anti-kink decouples, leaving a single CCGZ kink or anti-kink. For this solution, self-consistency 
requires a relation between the parameter $y$ and 
the fermion filling fraction of the valence bound states \cite{Dashen:1974ci}. This means that the physical size ($\sim c_0$) of the baryon is 
directly related to the number of valence fermions that
it binds, and results in intricate fermion dynamics during the scattering of DHN baryons \cite{Dunne:2011wu}. This static baryon
 solution can also
be boosted to a given velocity. In this paper we present the apparently new result that the DHN 
baryon can be expressed as a bound pair of twisted kinks, where the twist parameters are directly related to the baryon parameter $y$: 
see below, Section \ref{sec:2poles}.

\item
{\bf Real DHN Breather for GN$_2$:}
DHN also found in the GN$_2$  model an exact time-dependent self-consistent HF solution that is periodic in time in its rest-frame 
(known as the ``breather") \cite{Dashen:1974ci}:
\begin{eqnarray}
&&\text{condensate:} \quad S(x, t) =  \frac{1+b(2-K^2) e^{Kx}  - 2 a e^{Kx} \cos(\Omega t)+ e^{2Kx}}{1 + 2b e^{Kx}  + 2 a e^{Kx} \cos (\Omega t)+
 e^{2Kx}}  \label{eq:dhn-breather} \\
&& \hskip 3cm \Omega = \frac{2}{\sqrt{1+\epsilon^2}}, \quad K=\epsilon \Omega, \quad a=\frac{\epsilon}{2} \sqrt{4b^2 - 4 - K^2b^2} \nonumber \\
&&\text{filling-fraction consistency condition:} \quad b = (1-\nu) \frac{\sqrt{1+\epsilon^2}}{1-(2/\pi)\arctan \epsilon}
\nonumber
\end{eqnarray}
The DHN breather has two parameters,  $\epsilon$ and $b$,  characterizing the frequency and the amplitude of its oscillation. 
The breather also requires a self-consistency relation
 between the valence fermion filling fractions and the breather parameters \cite{Dashen:1974ci,Fitzner:2012kb}.

\end{enumerate}

\subsection{Building multiple-object solutions}

The aforementioned exact solutions have been generalized in various ways. 
First, as mentioned already, it is clear that  each can be boosted from its rest-frame. What is less clear is that they can be boosted 
independently, to describe scattering processes of 
independent objects. We show in this paper how this can be done in a fully self-consistent manner.

\begin{enumerate}

\item
The real CCGZ kinks for GN$_2$ can be combined into static multi-kink solutions \cite{Feinberg:2003qz}, and also kink-antikink crystals  
\cite{Thies:2006ti}. Exact solutions can also be 
given describing the scattering of arbitrary combinations of kinks and anti-kinks, with arbitrary velocities. This construction is based on 
the fact that the logarithm of the scalar condensate $S$ 
satisfies the Sinh-Gordon (ShG) equation \cite{Klotzek:2010gp,Fitzner:2010nv}, so these solutions can be constructed from the 
corresponding ShG solitons \cite{Jevicki:2009uz}.  
No fermion filling-fraction self-consistency condition is required.

\item
Takahashi et al \cite{Takahashi:2012pk} have recently presented an algebraic construction for  {\it static} multi-twisted-kink solutions
for the NJL$_2$ model, and twisted crystalline solutions 
were constructed in  \cite{Basar:2008ki}.  In this present paper we give new results for the {\it time-dependent} scattering of arbitrary 
combinations of twisted kinks, with arbitrary velocities. 
Note that the twisted kinks {\it do not} satisfy the Sinh-Gordon equation, so the construction uses other methods. We find a simple 
closed-form solution as a ratio of determinants, for both 
the static and time-dependent multi-twisted-kink solutions.

\item
The scattering of two DHN baryons for the GN$_2$ model was solved in \cite{Dunne:2011wu}, and an algorithmic procedure for the
description of multi-DHN-baryon scattering was presented 
in \cite{Fitzner:2012gg}. In this paper we show that DHN baryons can be constructed as bound twisted kinks, and therefore the 
scattering of DHN baryons can be described as special cases 
of the scattering of twisted kinks, for which we have a closed-form solution.

\item
Our construction leads to two new results concerning breathers. First, we find twisted breather solutions for the NJL$_2$ model, and 
we find solutions describing the scattering of any 
number of these twisted breathers. Second, as a consequence, we find the general solution for the scattering of any number of GN$_2$ 
breathers. This is consistent with the partial results 
of \cite{Fitzner:2012kb}. Indeed, our general construction describes the scattering of any number of any of these objects: real kinks, 
twisted kinks, DHN GN$_2$ baryons and breathers, and 
NJL$_2$ breathers.

\end{enumerate}

\subsection{Dirac equation and  kinematic notation}

\begin{figure}[htb]
\includegraphics[scale=0.4]{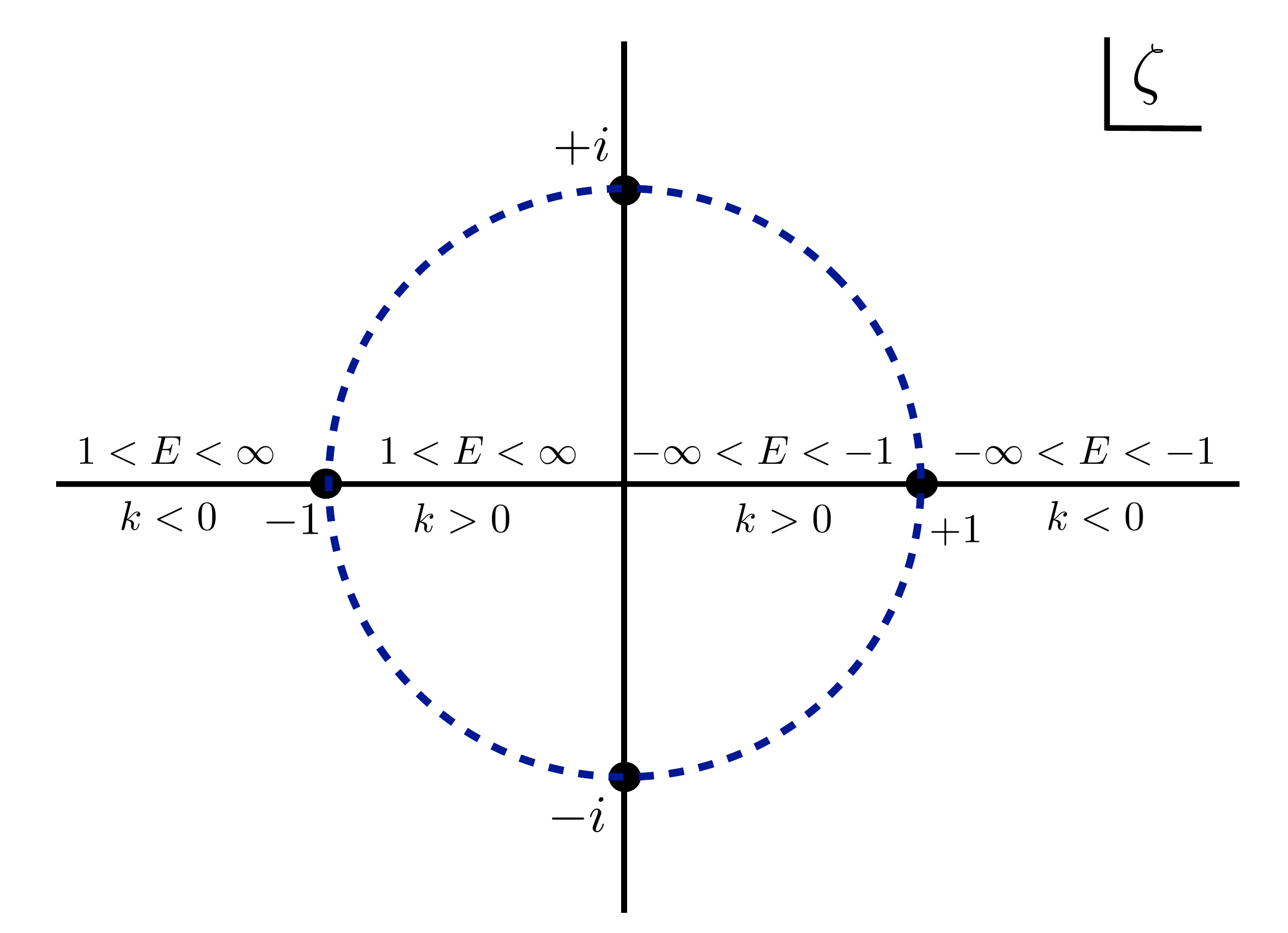}
\caption{The spectral $\zeta$ plane, indicating the regions of positive and negative energy and momentum. We have set the mass 
scale $m=1$. Note that for $\zeta$ outside the unit 
circle the boost has a positive velocity, negative inside the unit circle. Bound states, having $|E|<1$, correspond to a $\zeta$ of 
magnitude 1, lying on the unit circle.}
\label{fig:spectral-plane}
\end{figure}

We consider the TDHF problem (\ref{njl-tdhf}) for the NJL$_2$ model, and later we specialize to solutions of the GN$_2$ model.
We work with the following representation of the Dirac matrices:
\begin{equation}
\gamma^0 = \sigma_1, \quad \gamma^1 = i \sigma_2, \quad \gamma_5 = \gamma^0 \gamma^1 = - \sigma_3
\label{1b}
\end{equation}
and it is convenient to adopt light-cone coordinates (note that $\bar z$ is not the complex conjugate of $z$):
\begin{equation}
z=x-t, \quad \bar{z} = x+t, \quad \partial_0 = \bar{\partial}-\partial, \quad \partial_1 = \bar{\partial} + \partial 
\label{1c}
\end{equation}
The energy $E$ and momentum $k$ can be written in terms of the light-cone spectral parameter $\zeta$:
\begin{equation}
k = \frac{1}{2}\left( \zeta- \frac{1}{\zeta} \right), \quad E = - \frac{1}{2} \left( \zeta + \frac{1}{\zeta} \right)
\label{1d}
\end{equation}
where we measure energies and momenta in units of $m$, the dynamically generated fermion mass. 
We have included a minus 
sign in the definition of $E$ since for the consistency 
condition we will be summing over negative energy states in the Dirac sea. The various regions of the spectral plane, 
with corresponding energy and momentum, are shown in Fig.~\ref{fig:spectral-plane}.

The boost parameter $\eta$, rapidity $\xi$ and velocity $v$ are related by:
\begin{equation}
\eta = e^{\xi} = \sqrt{\frac{1+v}{1-v}}, \quad v = \frac{\eta^2 -1}{\eta^2+1}
\label{1e}
\end{equation}
Under a Lorentz boost, the light-cone variables transform as:
\begin{equation}
z \to \eta z, \quad \bar{z} \to \eta^{-1} z, \quad \zeta \to \eta \zeta
\label{1f}
\end{equation}
and the Lorentz scalar argument of a plane wave is written as:
\begin{equation}
k_{\mu} x^{\mu} = - \frac{1}{2} \left( \zeta \bar{z}- \frac{z}{\zeta} \right)
\label{1g}
\end{equation}
In terms of these variables, and in terms of the complex condensate (\ref{eq:delta}), the Dirac equation for the two-component 
spinor  ($\psi_1=\psi_L, \psi_2 = \psi_R$) reads:
\begin{equation}
2i \bar{\partial}\psi_2 = \Delta \psi_1\quad , \quad 2i \partial \psi_1 = -\Delta^* \psi_2
\label{eq:dirac}
\end{equation}

\section{General TDHF solution}
\label{sec:tdhf}
\subsection{Transparent potential}
\label{sec:transparent}
In a recent paper, a large class of transparent, time-dependent, scalar-pseudoscalar Dirac potentials has been constructed 
\cite{Dunne:2013tr}.
The method used was a generalization of the method invented by Kay and Moses for finding all static, transparent Schr\"odinger
potentials \cite{kaymoses}. We collect the main results, referring to Ref.~\cite{Dunne:2013tr} for proofs and more details.
We make the following ansatz for the continuum spinor
\begin{equation}
\psi_{\zeta} = \frac{1}{\sqrt{1+\zeta^2}} \left( \begin{array}{c} \zeta \chi_1 \\ - \chi_2 \end{array} \right) e^{i(\zeta \bar{z}-z/\zeta)/2},
\label{T1}
\end{equation}
where $\chi_1$ and $\chi_2$ approach some constant for $x \to \infty$.
In that case, the continuum 
spinor behaves like a plane wave travelling to the right for $x\to -\infty$, as well as for $x\to \infty$ (for $k>0$); hence it is manifestly
reflectionless. 

The basic ingredients in the construction of $\Delta$ and $\psi_{\zeta}$ are $N$  
``plane wave" factors $e_n, f_n$, with complex spectral parameters $\zeta_n$,
\begin{equation}
e_n = e^{i(\zeta_n^* \bar{z} - z/\zeta_n^*)/2}, \quad f_n = \frac{e_n}{\zeta_n^*}, \quad n=1,...,N
\label{T2}
\end{equation}
$N$ is the number of bound states. The reduced spinor components $\chi_{1,2}$ in (\ref{T1}) are written as finite sums with $N$ poles:
\begin{eqnarray}
\chi_1 & = &    1 + i \sum_{n=1}^N \frac{1}{\zeta-\zeta_n} e_n^* \phi_{1,n} ,
\nonumber \\
\chi_2 & = &    1 - i \sum_{n=1}^N \frac{\zeta}{\zeta-\zeta_n} e_n^* \phi_{2,n}, 
\label{T3}
\end{eqnarray}
Here $\phi_{1,n}$ and $\phi_{2,n}$ are 2$N$ functions defined as the solutions of the following systems of {\it linear, algebraic} 
equations,
\begin{eqnarray}
\sum_{m=1}^N \left( \omega+B\right)_{nm}\phi_{1,m} & = &  e_n,
\nonumber \\
\sum_{m=1}^N \left( \omega+B\right)_{nm}\phi_{2,m} & = & - f_n.
\label{T4}
\end{eqnarray}
Here, $\omega$ is a constant, hermitean but otherwise arbitrary $N\times N$ matrix, and $B$ is an $N\times N$ matrix constructed
from the basis functions, $e_n(z, \bar z)$, and spectral parameters, $\zeta_n$, as follows:
\begin{equation}
B_{nm} = i \frac{e_n e_m^*}{\zeta_m-\zeta_n^*}.
\label{T5}
\end{equation}
The $\zeta_n$ can be identified with the positions of the bound state poles of $\psi_{\zeta}$ in the complex $\zeta$-plane, see Eq.~(\ref{T3}).
To simplify the notation, we denote by $e, f, \phi_1, \phi_2$ the $N$-dimensional vectors with components $e_n,f_n,\phi_{1,n},\phi_{2,n}$,
respectively, whereas $\omega$ and $B$ denote $N \times N$ matrices. Eq.~(\ref{T4}) becomes
\begin{eqnarray}
(\omega+B) \phi_1 & =  & e,
\nonumber \\
(\omega + B) \phi_2 & = & - f.
\label{T6}
\end{eqnarray}
As shown in Ref.~\cite{Dunne:2013tr}, the $\phi_n$ are $N$ bound state spinors and $\psi_{\zeta}$ is the continuum spinor belonging to the 
transparent Dirac potential 
\begin{equation}
\Delta  =   1 - i e^{\dagger} \phi_2 = 1 + i \phi_1^{\dagger} f = 1+i e^{\dagger} \frac{1}{\omega+B} f .
\label{T7}
\end{equation} 
The three different expressions for $\Delta$ given here are equivalent owing to Eq.~(\ref{T6}).
Let us introduce a 3rd vector $g$ in addition to $e,f$ defined in Eq.~(\ref{T2}), with components
\begin{equation}
g_n = \frac{e_n}{\zeta-\zeta_n^*}.
\label{T8}
\end{equation}
This yields more compact expressions for $\chi_1, \chi_2$ as well, 
\begin{eqnarray}
\chi_1  & = &   1 + i g^{\dagger} \phi_1,
\nonumber \\
\chi_2 & = &   1 - i \zeta g^{\dagger} \phi_2.
\label{T9}
\end{eqnarray}
Furthermore,  simple expressions in terms of determinants have been presented in  \cite{Dunne:2013xta,Dunne:2013tr} for the 
condensate $\Delta$ and the spinor components $\chi_{1,2}$.

The bound state spinors $\phi_n$ are in general neither orthogonal nor normalized. A set of properly orthonormalized spinors
can be constructed via 
\begin{equation}
\hat{\phi}_n = \sum_{m=1}^N C_{nm} \phi_m, \quad \int_{-\infty}^{\infty} dx \hat{\phi}_n^{\dagger} \hat{\phi}_m = \delta_{n,m}.
\label{T10}
\end{equation}
As shown in \cite{Dunne:2013tr}, the matrix $C$ then satisfies the condition,
\begin{equation}
2 C \omega^{-1} C^{\dagger} = 1.
\label{T11}
\end{equation}
This was derived under the assumption that ${\rm Im}\, k_n>0 $, where 
\begin{equation}
k_n = \frac{1}{2} \left( \zeta_n - \frac{1}{\zeta_n} \right)
\label{T12}
\end{equation}
is the complex momentum belonging to the $n$-th bound state.
The following asymptotic behavior of the potential was found in \cite{Dunne:2013tr},
\begin{equation}
\lim_{x\to - \infty}\Delta =  1 , \quad \lim_{x \to \infty}\Delta  =  \prod_{n=1}^N \frac{\zeta_n}{\zeta_n^*}= e^{i\Theta}.
\label{T13}
\end{equation}
This shows that $\Delta$ has a chiral twist $e^{i\Theta}$ where the chiral twist angle $\Theta$ can be computed
by simply adding up the phases of all bound state pole parameters $\zeta_n$,
\begin{equation}
\Theta = 2 \sum_{n=1}^N \theta_n, \qquad \zeta_n = |\zeta_n| e^{i\theta_n}.
\label{T14}
\end{equation}
The spinor components have the asymptotic behavior
\begin{eqnarray}
\lim_{x\to - \infty}\chi_1 & = &   1 , \quad \lim_{x \to \infty}\chi_1  =  \prod_{n=1}^N \frac{\zeta-\zeta_n^*}{\zeta-\zeta_n},
\label{T15} \\
\lim_{x\to - \infty}\chi_2 & = &   1 , \quad \lim_{x \to \infty}\chi_2  = \prod_{n=1}^N \frac{\zeta_n}{\zeta_n^*}  \frac{\zeta-\zeta_n^*}{\zeta-\zeta_n}.
\label{T16}
\end{eqnarray} 
From Eq.~(\ref{T15}) we can read off the fully factorized, unitary transmission amplitude $T(\zeta)$ with the expected 
pole structure,
\begin{equation}
T(\zeta) =  \prod_{n=1}^N \frac{\zeta-\zeta_n^*}{\zeta-\zeta_n}, \qquad |T(\zeta)|=1.
\label{T17}
\end{equation}
The extra factors in the product in (\ref{T16}) are due to the chiral twist of the potential $\Delta$ which 
also affects the spinors. 

\subsection{Self-consistency}
\label{sec:selfco}
We now show that this solution also gives a self-consistent solution to the fully quantized TDHF problem (\ref{njl-tdhf}), 
provided certain filling-fraction conditions 
are satisfied by the combined soliton-fermion system, generalizing the conditions already found by DHN and Shei 
\cite{Dashen:1974ci,Shei:1976mn}.
The TDHF potential $\Delta$ receives contributions from the Dirac sea and the valence bound
states,
\begin{equation}
\Delta  =  -2Ng^2 \left( \langle \psi_1^*\psi_2 \rangle_{\rm sea} + \langle \psi_1^* \psi_2 \rangle_{\rm b} \right),
\label{T18}
\end{equation}
with
\begin{eqnarray}
\langle \psi_1^*\psi_2 \rangle_{\rm sea} & = & - \frac{1}{2} \int_{1/\Lambda}^{\Lambda} \frac{d\zeta}{2\pi} \frac{1}{\zeta} \chi_1^* \chi_2,
\label{T19} \\
\langle \psi_1^* \psi_2 \rangle_{\rm b} & = & \sum_n \nu_n \hat{\phi}_{1,n}^* \hat{\phi}_{2,n}.
\label{T20}
\end{eqnarray}
The integration limits in (\ref{T19}) correspond to a symmetric momentum cutoff $\pm\Lambda/2$ in ordinary coordinates.
We insert the expressions for $\chi_1, \chi_2$ from (\ref{T9}) and isolate the $\zeta$-dependence of the
integrand in the continuum part (\ref{T19}). The integrand contains only simple poles in the complex $\zeta$-plane, so that the integration over
$d\zeta$ with a cutoff can easily be performed. The pole at $\zeta=0$ yields the divergent contribution
\begin{equation}
\left. \langle \psi_1^*\psi_2 \rangle_{\rm sea}\right|_{\rm div} =  - \frac{\Delta}{2\pi} \ln \Lambda.
\label{T21}
\end{equation}
If one inserts this into (\ref{T18}) and uses the vacuum gap equation
\begin{equation}
\frac{Ng^2}{\pi} \ln \Lambda = 1,
\label{T22}
\end{equation}
one finds that this part gives self-consistency by itself. Requiring that   
the convergent part of the sea contribution cancels the bound state contribution should give us the relationship
between the bound state occupation fractions $\nu_n$ and the parameters of the solution, provided the solution is 
self-consistent. The computation of the convergent part  of the sea contribution is straightforward. To present the 
result in a concise form, we introduce a diagonal matrix $M$,
\begin{equation}
M_{nm} = -i \delta_{nm} \ln(-\zeta_n^*).
\label{T23}
\end{equation} 
(Logarithms of $\zeta_n$ appear if one integrates over $d\zeta$, as a result of the simple poles in the complex $\zeta$ plane.)
The convergent part of (\ref{T19}) can then be simplified to
\begin{equation}
\left. \langle \psi_1^*\psi_2 \rangle_{\rm sea}\right|_{\rm conv} = - \frac{1}{4\pi} \phi_1^{\dagger}  \left( \omega M^{\dagger} +
M \omega \right) \phi_2  
\label{T24}
\end{equation}
The bound state contribution (\ref{T20}) is evaluated with the help of Eq.~(\ref{T11}). After introducing another diagonal matrix $N$,
\begin{equation}
N_{nm} = 4 \pi \delta_{nm} \nu_n,
\label{T25}
\end{equation}
it can be written as
\begin{equation}
\langle \psi_1^* \psi_2 \rangle_{\rm b} =   \frac{1}{4\pi} \phi_1^{\dagger} \left( C^{\dagger} N C  \right)  \phi_2.
\label{T26}
\end{equation}
Expressions (\ref{T24}) and (\ref{T26}) cancel
if we require that 
\begin{equation}
\omega M^{\dagger} + M \omega = C^{\dagger} N C.
\label{T27}
\end{equation}
This is the self-consistency relation determining the bound state occupation fractions. It can be cast into a more convenient form by
combining Eqs.~(\ref{T11}) and (\ref{T27}) as follows. From our experience with concrete applications of this formalism, it appears that 
$\omega$ should be chosen as a positive definite matrix to avoid singularities in $\Delta$ as a function of $(x,t)$. Assuming that $\omega$
is positive definite, it has the unique Cholesky decomposition
\begin{equation}
\omega = L L^{\dagger}
\label{T28}
\end{equation}
where $L$ is a lower triangular matrix. From (\ref{T11}) we conclude that the matrix
\begin{equation}
V= \sqrt{2} C \frac{1}{L^{\dagger}}  
\label{T29}
\end{equation}
is unitary. The self-consistency condition (\ref{T27}) can then be transformed into the final form
\begin{equation}
2 \left(L^{\dagger} M^{\dagger} \frac{1}{L^{\dagger}} + \frac{1}{L} M L \right) = V^{\dagger} N V.
\label{T30}
\end{equation}
Thus, the eigenvalues of the matrix
on the left hand side of (\ref{T30}) determine the diagonal entries of the matrix $N$, which yield the fermion filling fractions $\nu_n$ in (\ref{T25}). 
To test whether a given candidate solution is self-consistent, one has to 
confirm that all eigenvalues are between 0 and $4 \pi$, thereby satisfying the self-consistency condition with physical occupation fractions $\nu_n \in [0,1]$. As an alternative to the Cholesky
decomposition, Eq.~(\ref{T30}) remains valid if one replaces $L$ by $\sqrt{\omega}$, which can
be computed by diagonalizing $\omega$ first.

\subsection{Vanishing fermion density}
\label{sec:density}
Due to strong constraints from chiral symmetry in 1+1 dimensions, the massless NJL$_2$ model does not allow any localized
fermion density or current \cite{Karbstein:2007}. Similarly, there is no localized energy or momentum density \cite{Brendel:2010}. This 
follows from the conservation laws 
\begin{equation}
\partial_{\mu} j_V^{\mu}  =   0, \qquad \partial_{\mu} j_A^{\mu} = 0 , \qquad
\partial_{\mu} {\cal T}^{\mu \nu} = 0
\label{T31}
\end{equation}
together with the fact that 
\begin{eqnarray}
j_V^0 & = &  j_A^1 = \psi^{\dagger}\psi, \qquad j_V^1 = j_A^0 = \psi^{\dagger} \gamma_5 \psi
\nonumber \\
{\cal T}^{00} & = &  {\cal T}^{11} = {\cal H}, \qquad {\cal T}^{01} = {\cal T}^{10} = {\cal P}
\label{T32}
\end{eqnarray}
in the massless NJL$_2$ model.
The conservation laws (\ref{T31}) remain valid in TDHF approximation. Since the bound states carry lumps of localized fermions, there must
be an exact cancellation between continuum states and bound states for all of these densities. 
As a consistency test of the above TDHF solution, let us check this cancellation explicitly for the simplest case, the fermion density
$\rho=j_V^0$. The induced fermion density in the Dirac sea is
\begin{equation}
\rho_{\rm ind} = \int_0^{\infty} \frac{d\zeta}{2\pi} \frac{\zeta^2+1}{2\zeta^2} \left( \psi_{\zeta}^{\dagger} \psi_{\zeta}-1 \right)
= \frac{1}{2} \int_0^{\infty} \frac{d\zeta}{2\pi} \left\{ \left| \chi_1 \right|^2 - 1 + \frac{1}{\zeta^2}\left( \left| 
\chi_2  \right|^2 - 1\right) \right\}.
\label{T33}
\end{equation}
If we replace $\chi_1,\chi_2$ by the expressions given in (\ref{T9}), we can simplify the result after some straightforward computations to 
\begin{equation}
\rho_{\rm ind} =  \partial_x \int_0^{\infty} \frac{d\zeta}{2\pi} g^{\dagger} \frac{1}{\omega+B} g.
\label{T34}
\end{equation}
Inserting the $g$'s and performing the integration over $d\zeta$, the result can be  written as
\begin{equation}
\rho_{\rm ind}  =  \frac{1}{2\pi} {\rm Tr} \left[ \left( \omega M^{\dagger} + M \omega \right) \partial_x \frac{1}{\omega+B}\right].
\label{T35}
\end{equation}
The density from the bound states with occupation fractions $\nu_n$ yields \cite{Dunne:2013tr}
\begin{equation}
\rho_b   =    \sum_n \nu_n \hat{\phi}_n^{\dagger} \hat{\phi}_n 
 =  - \frac{1}{2\pi} {\rm Tr} \left[ \left(C^{\dagger} N C\right) \partial_x \frac{1}{\omega+B} \right]
\label{T36}
\end{equation}
If the self-consistency condition (\ref{T27}) is satisfied, the bound state density (\ref{T36}), and the induced fermion density
in the sea (\ref{T35}) cancel exactly. The vanishing of the current density $j_V^1$ can be proven in a similar 
manner, the only difference being that $\partial_x$ gets replaced by $\partial_t$ everywhere. 

\subsection{Time delays and masses}
\label{sec:masses}
In standard soliton theory, the outcome of a scattering process is expressed via the time delay experienced by the solitons
during the collision. As already discussed in \cite{Fitzner:2012gg}, the situation is more complicated if multi-soliton bound states
are involved. In this case the shape of the bound state may be affected as well. In the present work, we face the additional complication
that the phases
entering the breather oscillation may be changed during the scattering process. The best way to define the outcome of such a scattering
process
of composite multi-soliton objects is to compare the potential $\Delta$ for a cluster of kinks moving with a common velocity $v_0$ before
and after the collision. Inspection of a few cases with small number of kinks shows the following general pattern: The change in $\Delta$ for
a cluster involving $K$ kinks consists of an overall twist factor $\tau$ and rescalings of all the elementary functions $e_n$ by complex numers 
$\lambda_n$, 
\begin{equation}
\Delta_{\rm out}(e_{i_1},...,e_{i_K}) = \tau \Delta_{\rm in}(\lambda_{i_1}e_{i_1},...,\lambda_{i_K}e_{i_K})
\label{T37}
\end{equation}
with
\begin{eqnarray}
\tau & = & \prod_{n(v_n<v_0)} \frac{\zeta_n}{\zeta_n^*} \prod_{m(v_m>v_0)}   \frac{\zeta_m^*}{\zeta_m} ,
\nonumber \\
\lambda_n & = & \prod_{m(v_m<v_0)}\left(\frac{\zeta_n^*-\zeta_m^*}{\zeta_n^*-\zeta_m}\right)  \prod_{k(v_k>v_0)}
\left(\frac{\zeta_n^*-\zeta_k}{\zeta_n^*-\zeta_k^*}\right) 
\label{T38}
\end{eqnarray}
Alternatively, one could interpret the rescalings of the $e_n$ as a modification of the matrix $\omega^0$ of the cluster (one block out
of the full, block-diagonal matrix $\omega$), 
\begin{equation}
\left. \omega_{nm}^0 \right|_{\rm out} = \frac{\left. \omega_{nm}^0 \right|_{\rm in}}{\lambda_n \lambda_m^*}.
\label{T39}
\end{equation}
The twist factor $\tau$ can readily be understood in terms of the chiral twists of the solitons involved in the scattering process.
The elementary factors entering the expression for $\lambda_n$ also have a simple interpretation. The transmission amplitude 
of a fermion with spectral parameter $\zeta$ scattering off soliton $m$ is 
\begin{equation}
T_m(\zeta) = \frac{\zeta-\zeta_m^*}{\zeta-\zeta_m}.
\label{T40}
\end{equation}
Hence the factor $\lambda_n$ in (\ref{T38}) can be expressed in terms of transmission amplitudes of a fermion on all solitons
not belonging to the cluster, evaluated at the complex spectral parameter $\zeta_n^*$, the complex conjugate of the bound state pole 
position,
\begin{equation}
\lambda_n =  \prod_{m(v_m<v_0)} T_m(\zeta_n^*)  \prod_{k(v_k>v_0)} \frac{1}{T_k(\zeta_n^*)}.
\label{T41}
\end{equation}

Another question of interest concerns the masses of clusters of solitons. 
In Ref.~\cite{Brendel:2010}, a formula for the mass of TDHF solutions of the NJL$_2$ model was derived. Starting from
Eqs.~(\ref{T31}, \ref{T32}) for the energy momentum tensor, it was found that the mass can be expressed in terms of the
asymptotic behavior of the fermion phase shift for $k\to \infty$, 
\begin{equation}
M = \frac{N}{\pi} \lim_{k\to \infty} k \delta(k).
\label{T42}
\end{equation}
Here, $\delta(k)$ is the phase of the (unimodular) fermion transmission amplitude $T(k)$.
For a single twisted kink, this reproduces the original result of Shei \cite{Shei:1976mn},
\begin{equation}
M_1 = \frac{N}{\pi} \sin \varphi_1, \qquad \zeta_1=-e^{-i\varphi_1}.
\label{T43}
\end{equation}
According to (\ref{T17}), the full transmission amplitude factorizes into fermion-kink transition amplitudes, hence the 
phase shifts are additive, as expected for integrable systems. This holds independently of whether the solitons form
static bound states or breathers. Consequently, the mass of any compound of $n$ solitons is just the sum of the masses 
of the constituents --- the binding energy vanishes. This is consistent with what has already been known for static
bound states since Ref.~\cite{Shei:1976mn}, but generalizes to the breather case as well.  

An interesting spinoff results if we apply these insights to real $\Delta$, i.e., TDHF solutions of the
GN model.  A two-kink bound state has the mass 
\begin{equation}
M_{\rm kink}(\varphi_1) + M_{\rm kink}(\pi-\varphi_1) = \frac{2N}{\pi} \sin \varphi_1.
\label{T44}
\end{equation}
This relates the mass of the DHN baryon (or breather, for that matter) to the mass of the Shei kink ($\sin \varphi_1$
is the parameter $y$ in DHN). This is perhaps the most conspicuous manifestation of the long overlooked fact that twisted
kinks are the (hidden) constituents of the DHN baryon.

\section{Explicit Examples}
\label{sec:examples}

In this Section we illustrate the general solution to the TDHF problem (\ref{T7}, \ref{T9}, \ref{T30}) with 
several examples. We classify the applications according to the number of bound states or, equivalently, the number
of poles of the continuum spinors in the complex $\zeta$-plane.

\subsection{General solution with one pole: twisted kink}
\label{sec:1pole}

\begin{figure}[htb]
\includegraphics[scale=0.4]{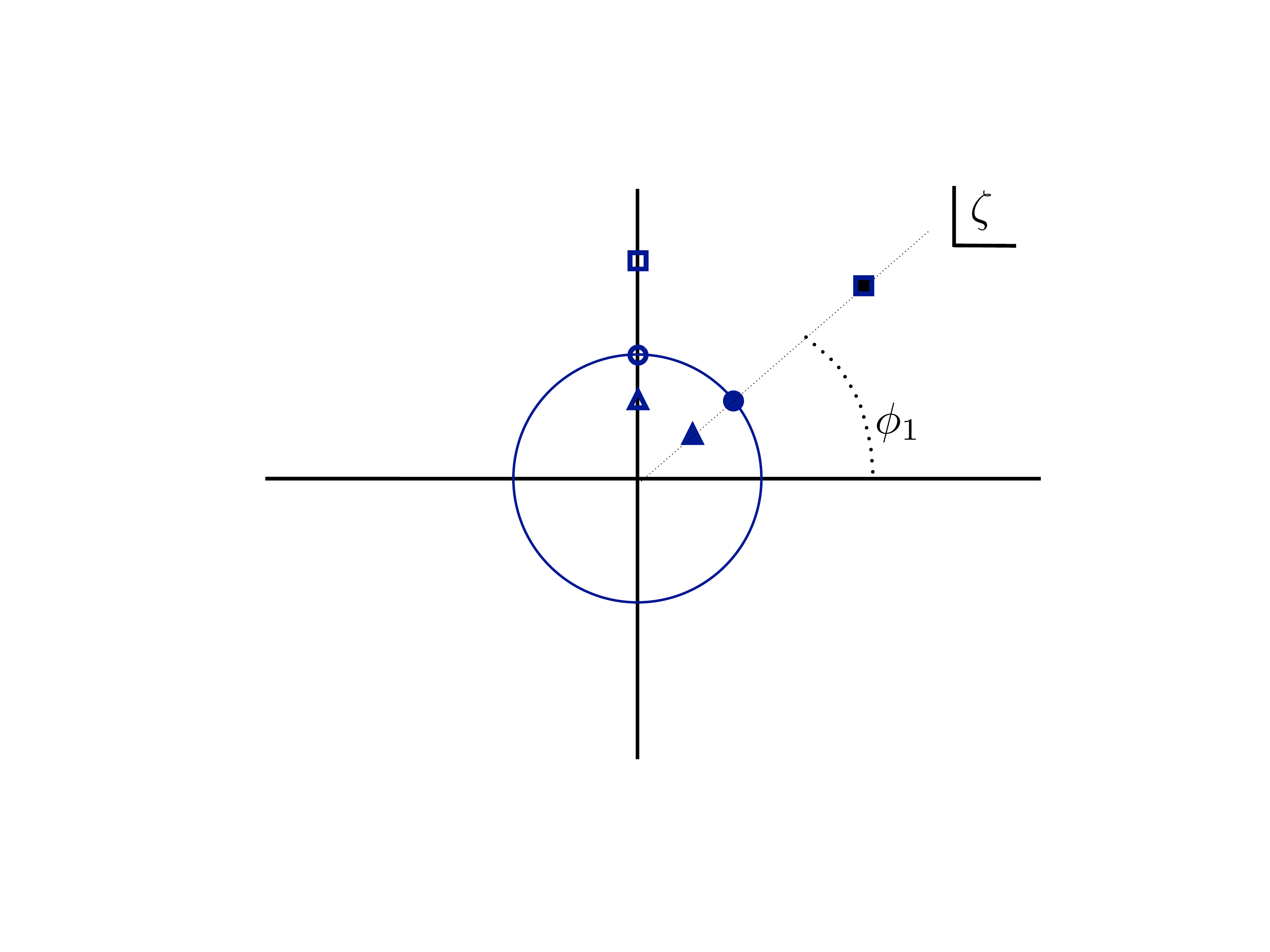}
\caption{A spectral $\zeta$ plane representation of single kinks. Each of the open square, circle and triangle on the positive imaginary 
axis represents a real GN$_2$ kink, with positive, zero or negative boost, with respect to the rest frame. Each of the full square, circle
and triangle on the ray at angle $\phi_1$ represents a complex twisted  NJL$_2$ 
kink with phase parameter $\phi_1$. Taken together, all these six points represent the scattering of six kinks, 3 of them real and 3 with
(equal) twist parameter $\phi_1$.}
\label{fig:spectral-plane-kinks}
\end{figure}

With one pole, the matrix $\omega$ is just a real number and the matrix $B$ reduces to a single function of $z,\bar{z}$. We parameterize
the position  $\zeta_1$ of the pole as
\begin{equation}
\zeta_1 = - \frac{e^{-i\varphi_1}}{\eta_1}, \quad \eta_1 = \sqrt{ \frac{1+v_1}{1-v_1} }.
\label{U1}
\end{equation}
The complex potential $\Delta$ can then be written as
\begin{equation}
\Delta = \frac{1 + e^{-2i \varphi_1} U_1}{1+ U_1}
\label{U2}
\end{equation}
with the real function
\begin{equation}
U_1 = \frac{B_{11}}{\omega_{11}} = \frac{\eta_1}{2 \omega_{11} \sin \varphi_1} \exp \left\{ \frac{\sin \varphi_1}{\eta_1} (\bar{z}+\eta_1^2 z) \right\}.
\label{U3}
\end{equation}
Expressed in ordinary coordinates, the argument of the exponential in $U_1$ reads
\begin{equation}
\frac{\sin \varphi_1}{\eta_1} (\bar{z}+\eta_1^2 z) = 2 \sin \varphi_1 \frac{x-v_1 t}{\sqrt{1-v_1^2}}.
\label{U4}
\end{equation}
This is the boosted form of the Shei twisted kink (\ref{eq:shei}) for the NJL$_2$ model. The role of the free parameter $\omega_{11}$ is to
shift the position of the kink. The phase and modulus of $\zeta_1$ are related to the chiral twist and the velocity of the kink, respectively, 
as illustrated in Fig. \ref{fig:spectral-plane-kinks}.
To cover the full range of chiral twists it is sufficient to restrict $\varphi_1$ to the interval $[0,\pi]$. In this case, $\sin \varphi_1 > 0$ and we 
have to choose 
$\omega_{11}>0$ in order to get a nonsingular $\Delta$. Notice that this definition of $\varphi_1$ also implies ${\rm Im}\, k_1>0$, as assumed 
in \cite{Dunne:2013tr}.
Turning to the self-consistency issue, the matrix $M$ introduced in (\ref{T23}) has
just one component: $M_{11}=-i\, 
\ln(-\zeta_1^*)=\varphi_1+i\,\ln \eta_1$. Thus, the NJL$_2$ filling fraction condition (\ref{T30}) gives:
\begin{eqnarray}
\nu_1=\frac{\varphi_1}{\pi}
\label{eq:n1-filling}
\end{eqnarray}
This self-consistent TDHF kink binds a number $n_v$ of valence fermions, where in the large $N_f$ limit the filling fraction $\nu_1=n_v/N_f$ 
is equal to the twist angle $\varphi_1$  divided by $\pi$.

We obtain the real kink solution (\ref{eq:ccgz}) of GN$_2$ by choosing $\varphi_1=\frac{\pi}{2}$ in (\ref{U2}, \ref{U3}). For GN$_2$ there is no 
filling fraction condition, as we do not have to impose a self-consistency condition on the pseudo scalar condensate.

\begin{figure}[htb]
\includegraphics[scale=0.5]{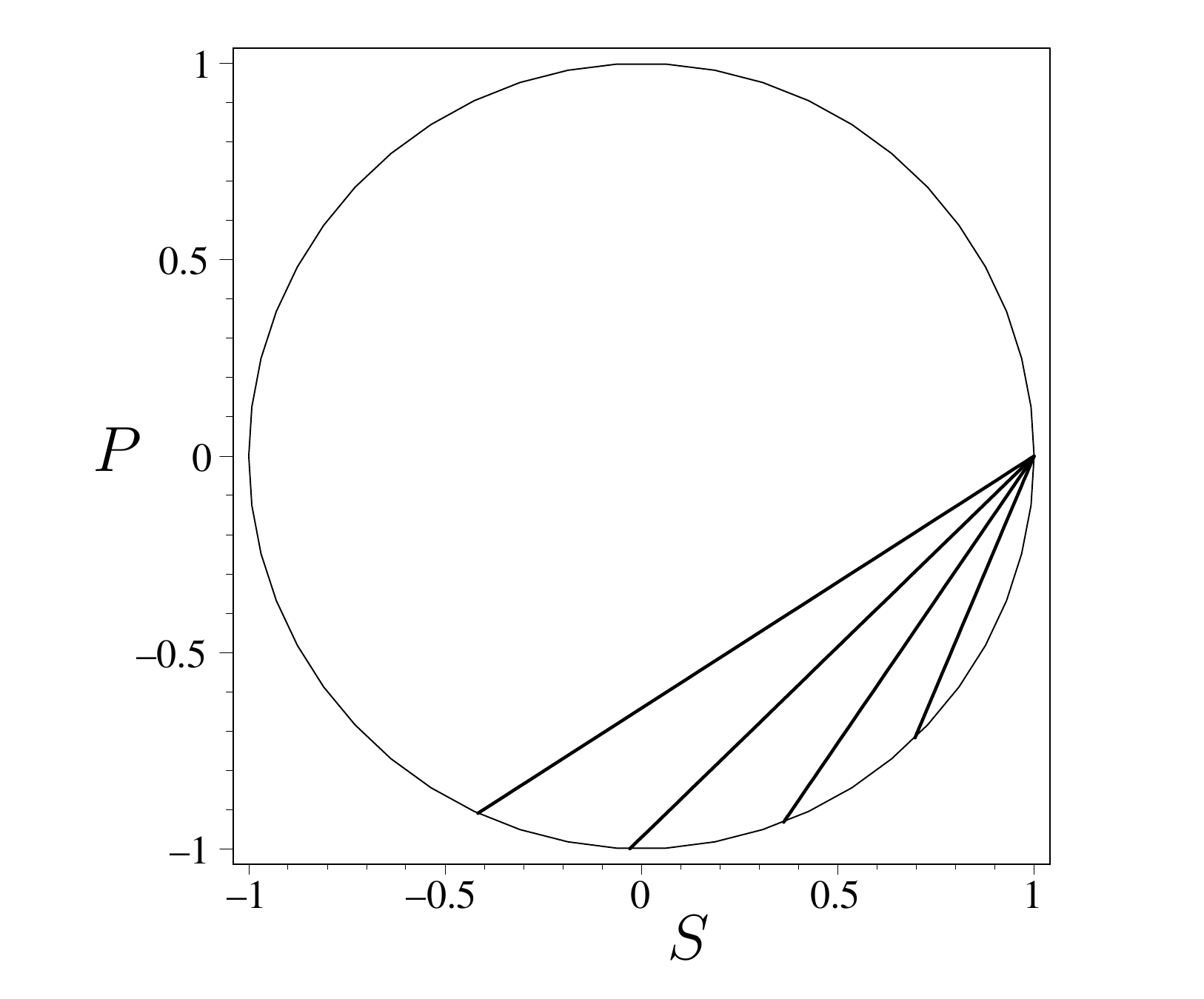}
\caption{($S,P$)-plot, of the scalar ($S$) and pseudoscalar ($P$) components of the condensate $\Delta$, for the four basic
twisted kinks used to build up most of the multi-kink configurations in this work, as explained in the text.}
\label{FIG0}
\end{figure}

In an $(S,P)$-plot, the twisted kink traces out a segment of a straight line, joining two points on the chiral circle. In our case, the starting point  
($x\to - \infty$) is always the point $(S=1,P=0)$, whereas the endpoint ($x\to \infty$) depends on the chiral twist. Most of the examples 
discussed below are based on constituent kinks with parameters $\varphi_1$ = 1.0, 0.8, 0.6, 0.4, shown in Fig.~\ref{FIG0} and, in  greater
detail, in the ancillary files to this paper (see {\tt 3dplot$\_$constituent$\_$kinks}).

\subsection{General solution with two poles: kinks, baryons and breathers} 
\label{sec:2poles}

With two poles, the matrices entering the general solution (\ref{T7}, \ref{T9}, \ref{T30}) are $2\times 2$. This enables us to work out
everything explicitly, including the self-consistency condition. The physics depends on the assumptions about the constant matrix $\omega$
and the pole positions $\zeta_{1,2}$. 

\subsubsection{Non-breather solutions} 
We find non-breather solutions by choosing a diagonal form of $\omega$ in (\ref{T7}). Introducing functions $U_i=B_{ii}/\omega_{ii}$
for $i=1,2$ in analogy to (\ref{U3}) and generalizing the parameterization (\ref{U1}) to $\zeta_{1,2}$, we find the potential 
\begin{equation}
\Delta = \frac{1+ e^{-2i \varphi_1} U_1 + e^{-2i \varphi_2} U_2 + b_{12} e^{-2i(\varphi_1+\varphi_2)} U_1 U_2}
{1+ U_1 + U_2 + b_{12} U_1 U_2}.
\label{U5}
\end{equation}
The interaction effects between the two twisted kinks are described by the real parameter $b_{12}$ given by
\begin{equation}
b_{12} = \left| \frac{\zeta_1 - \zeta_2}{\zeta_1-\zeta_2^*} \right|^2 = \frac{\eta_1^2+ \eta_2^2- 2\eta_1 \eta_2 \cos(\varphi_1-\varphi_2)}
{\eta_1^2+ \eta_2^2- 2\eta_1 \eta_2 \cos(\varphi_1+\varphi_2)}.
\label{U6}
\end{equation}
This is the $N=2$ case of a general formula valid for diagonal $\omega$, presented in Sect.~IIIB of \cite{Dunne:2013tr}.
Eq.~(\ref{U5}) gives the self-consistent potential for the scattering of two twisted kinks, with twist angles 
$\varphi_1$ and $\varphi_2$, and boost parameters $\eta_1$ and $\eta_2$, or for a bound state if one chooses $\eta_1=\eta_2$.
The filling-fraction consistency condition is simple when $\omega$ is diagonal. The $M$ matrix is
$M={\rm diag}(\varphi_1+i\, \ln \eta_1, \varphi_2+i\, \ln \eta_2)$. Thus, we find filling fractions
\begin{eqnarray}
\nu_1=\frac{\varphi_1}{\pi}\quad, \quad \nu_2=\frac{\varphi_2}{\pi},
\label{eq:n2-filling}
\end{eqnarray}
as expected from the asymptotics of the scattering problem.
If we are interested in solutions of the NJL$_2$ model with real $\Delta$, we are restricted to fermion number 0. In that case 
the self-consistency condition yields
\begin{equation}
\nu_1= \frac{\varphi_1}{\pi}, \quad \nu_2 = \frac{\varphi_2}{\pi} = \frac{\pi - \varphi_1}{\pi} = 1  - \nu_1
\label{U7}
\end{equation}
This corresponds to an ``exciton" in condensed matter language. 
In the GN case, we cannot take over the derivation of the self-consistency condition which was only valid for generic parameters. 
Now, the contributions of the 2 bound states give
 equal and opposite contributions to the condensate $\bar{\psi}{\psi}$, so that only the difference of the 
corresponding two equations of the NJL$_2$ model survives,
\begin{equation}
\nu_1 - \nu_2 = \frac{\varphi_1 - \varphi_2}{\pi} = \frac{2\varphi_1}{\pi} - 1
\label{U8}
\end{equation}
The baryon state of lowest energy for given baryon number has fully occupied negative energy bound state, 
corresponding to $\nu_2=1, \nu_1 = 2 \varphi_1/\pi$. This is the relation familiar from DHN.
\begin{figure}[htb]
\includegraphics[scale=0.4]{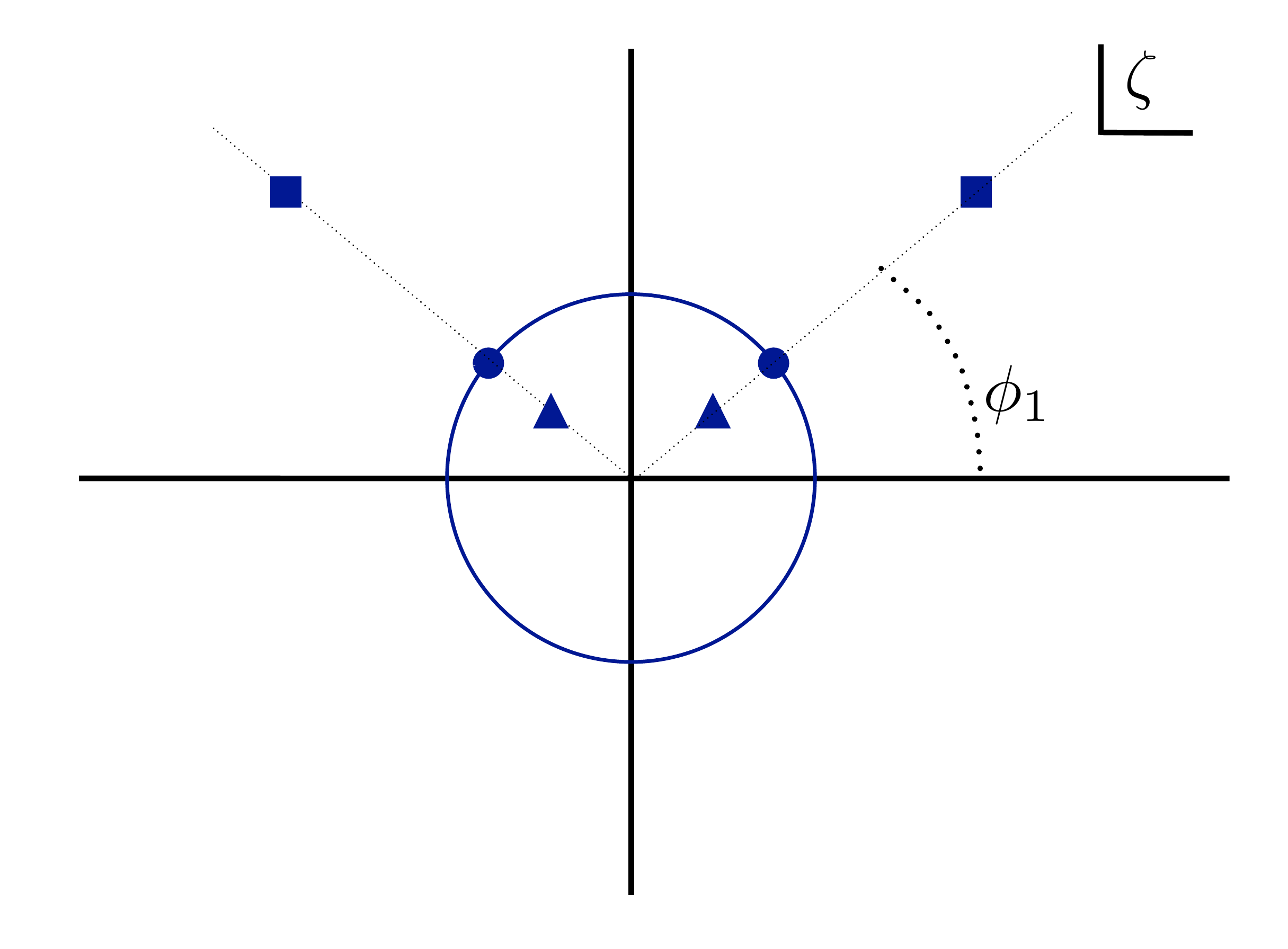}
\caption{A spectral $\zeta$ plane representation of  a real GN$_2$ baryon. The baryon is composed of two twisted kinks, one with chiral
angle $\phi_1$, and the other with $\pi-\phi_1$. The triangles (squares) have $\zeta_1<1$  ($\zeta_1>1$), corresponding to negative
(positive) boost parameter, while the circles correspond to a baryon at rest.}
\label{fig:spectral-plane-baryons}
\end{figure}

We can consider various special cases:
\begin{enumerate}
\item
Scattering of two GN$_2$ kinks. We obtain real kink solutions by setting $\varphi_1=\varphi_2=\frac{\pi}{2}$. Then 
\begin{equation}
S = \frac{1-U_1-U_2 + b_{12} U_1U_2}{1+U_1+U_2+b_{12} U_1U_2}, \quad b_{12} = \left( \frac{\eta_1-\eta_2}{\eta_1+\eta_2}\right)^2,
\label{U9}
\end{equation}
and $\sin \varphi_{1,2}=1$ in the definition of $U_{1,2}$.
This agrees with the $n=2$ case of the  general formula in \cite{Fitzner:2010nv}. There is no filling-fraction consistency condition.
\item
GN$_2$ baryon. We can also obtain a real solution by choosing $\varphi_2=\pi-\varphi_1$, together with $\omega_{11}=\omega_{22}$. 
To obtain a baryon we also choose $\eta_1=\eta_2$. Then $U_1=U_2$ and we find
\begin{equation}
S = \frac{1+ 2 \cos (2 \varphi_1) U_1 + \cos^2 \varphi_1 U_1^2}{1+ 2 U_1 + \cos^2 \varphi_1 U_1^2}
\label{U10}
\end{equation}
which agrees with the GN$_2$ baryon in (\ref{eq:dhn-baryon}). Thus, we see that the DHN GN$_2$ baryon is in fact a bound pair of
two twisted kinks, as depicted in Fig \ref{fig:spectral-plane-baryons}. The fermion filling-fractions are $\nu_2=1, \nu_1 = 2 \varphi_1/\pi$, as 
in DHN. Note that DHN have written the parameter $y$
which defines the size of the baryon in the form $y=\sin \theta$, without geometrical interpretation of the angle $\theta$. Now we 
see that $\theta$
is nothing but the angle $\varphi_1$ related to the twist of the constituent kinks. These constituents are well hidden inside the baryon, since 
the individual twisted kinks are not solutions of the GN model. The only observable which hints at this compositeness is the factorized fermion
transmission amplitude.  
\end{enumerate}

\begin{figure}[htb]
\includegraphics[scale=0.45]{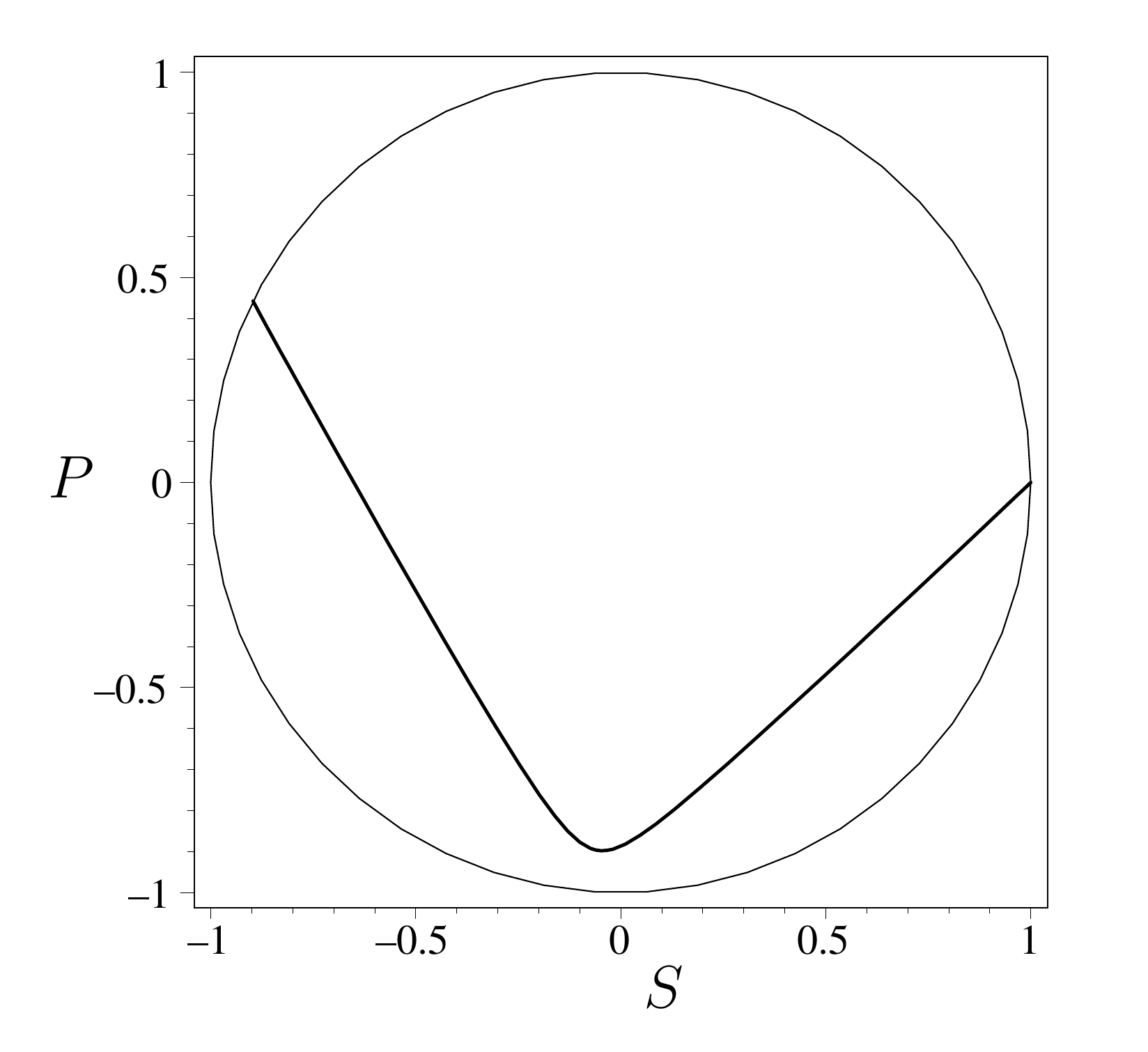}
\caption{($S,P$)-plot of a 2-kink, a bound state of two twisted kinks.}
\label{FIG1}
\end{figure}
\begin{figure}[htb]
\includegraphics[scale=0.5]{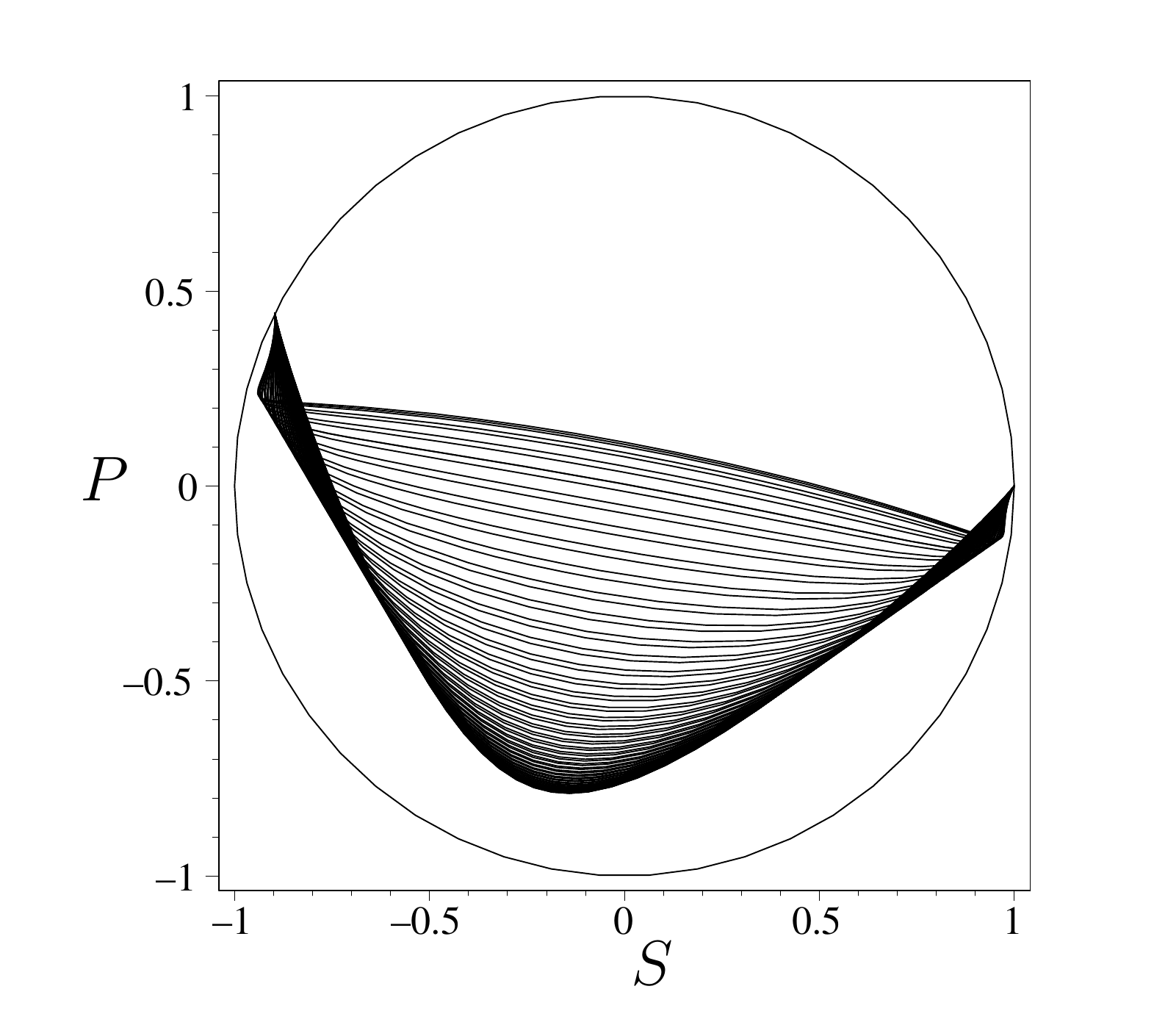}
\caption{($S,P$)-plot of a 2-breather made out of two kinks with the same parameters as the bound state in Fig.~\ref{FIG1}. The 
different curves illustrate the time dependence of the twisted breather, in equal time steps.}
\label{FIG2}
\end{figure}
\begin{figure}[htb]
\includegraphics[scale=0.5]{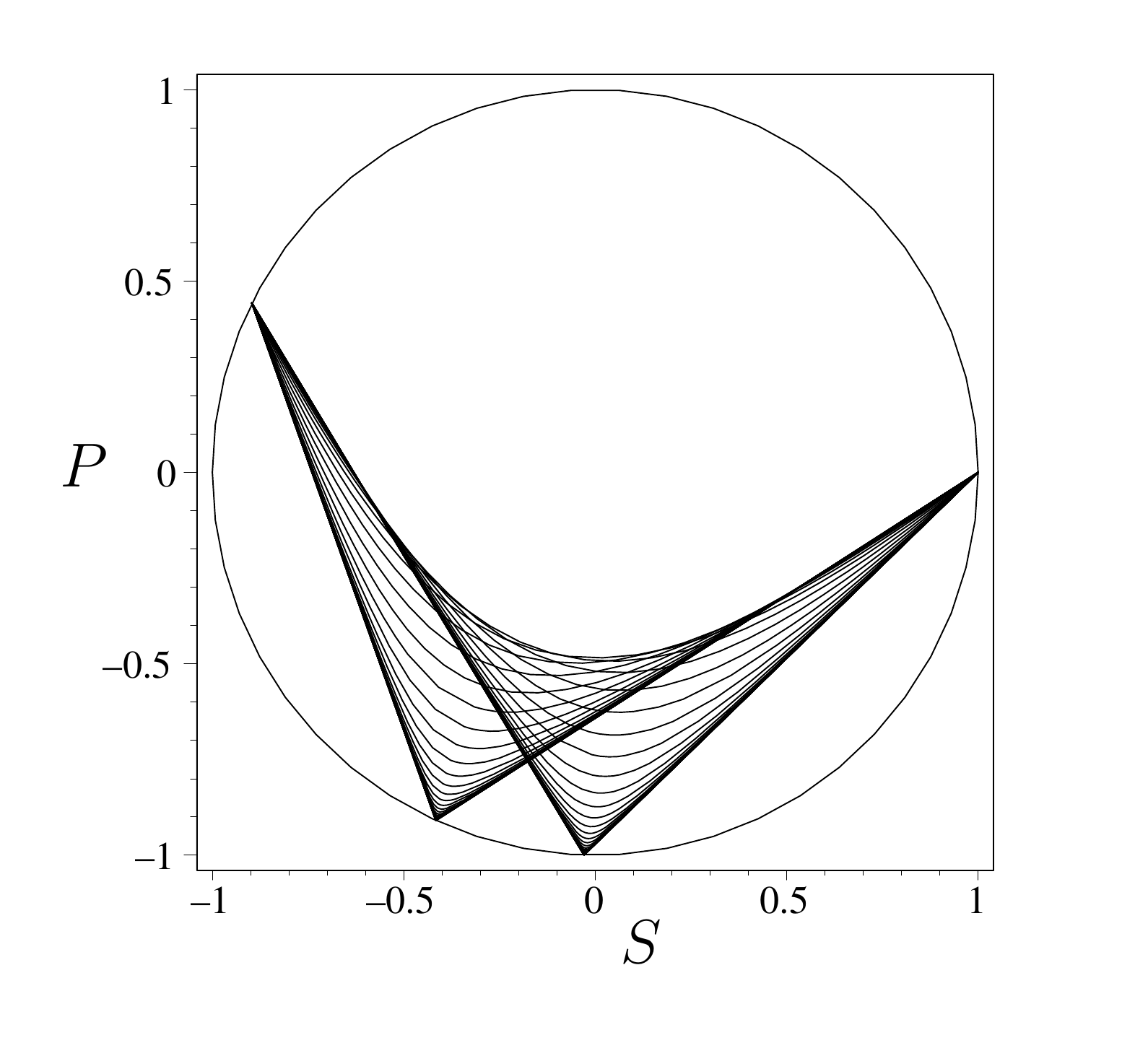}
\caption{($S,P$)-plot of the scattering process of two twisted kinks. The curves show the time dependence, in equal time steps. The initial and 
final states are open polygons with 2 segments, ending on the chiral circle.}
\label{FIG3}
\end{figure}

\subsubsection{Breather Solutions}

Breather solutions in the rest frame are obtained by choosing $\eta_1=\eta_2=1$ and a non-diagonal $2\times 2$ matrix $\omega$ 
in (\ref{T14}, \ref{T30}). Using the freedom of making translations in $x$ and $t$, we choose the following positive definite hermitean matrix:
\begin{eqnarray}
\omega = 
\begin{pmatrix}
{ \sec \chi & \tan \chi \cr
\tan \chi & \sec \chi
}
\end{pmatrix}
\label{eq:breather-omega}
\end{eqnarray}
Then we find for the GN$_2$ system where $\varphi_2=\pi-\varphi_1$:
\begin{eqnarray}
S & = & \frac{\cal N}{\cal D} \nonumber \\
{\cal N} & = & 1 + \frac{\cos(2\varphi_1)}{\sin \varphi_1 \cos \chi} e^{2 x \sin \varphi_1}+ \tan \chi e^{2 x \sin \varphi_1}
 \sin( 2 t \cos \varphi_1 + \varphi_1) + \frac{1}{4} \cot^2 \varphi_1 e^{4x \sin \varphi_1} \nonumber \\
{\cal D} & = & 1 + \frac{1}{\sin \varphi_1 \cos \chi} e^{2 x \sin \varphi_1} - \tan \chi e^{2 x \sin \varphi_1} \sin( 2 t \cos \varphi_1 + \varphi_1)
+ \frac{1}{4} \cot^2 \varphi_1 e^{4x \sin \varphi_1} 
\label{eq:n2-breather1}
\end{eqnarray}
This agrees (modulo translations in $x$ and $t$) with the DHN GN$_2$ breather (\ref{eq:dhn-breather}) if we use the following identifications,
\begin{eqnarray}
 \epsilon = \tan \varphi_1, \qquad b= \frac{1}{\cos \varphi_1 \cos \chi}, \qquad a = \tan \varphi_1 \tan \chi .
\label{U11}
\end{eqnarray}
The limit $\chi\to 0$ of (\ref{eq:n2-breather1}) yields back the static DHN baryon (\ref{U10}) up to a shift in $x$, as can be seen by
setting
\begin{equation}
U_1 = \frac{e^{2x\sin \varphi_1}}{2 \sin \varphi_1}.
\label{U12}
\end{equation}

A new twisted breather for the NJL$_2$ model is obtained by choosing the off-diagonal mixing matrix (\ref{eq:breather-omega}), and 
relaxing the reality condition  (so that $\varphi_2\neq \pi-\varphi_1$) on the twist angles. This is the most complicated TDHF solution with 
two poles. In order to exhibit its structure, we first write down the potential $\Delta$ in the form
\newpage
\begin{eqnarray}
\Delta & = & \frac{\cal N}{\cal D}
\nonumber \\
{\cal N} & = & 1 + \frac{1}{\cos \chi} \left( \frac{\zeta_1}{\zeta_1^*}U_1+\frac{\zeta_2}{\zeta_2^*}U_2 \right) + b_{12}\frac{\zeta_1 \zeta_2}
{\zeta_1^*\zeta_2^*} U_1 U_2 - \tan \chi \left( \frac{\zeta_2}{\zeta_1^*}B_{12} + \frac{\zeta_1}{\zeta_2^*} B_{21} \right)
\nonumber \\
{\cal D} & = &  1 + \frac{1}{\cos \chi} \left(U_1+ U_2 \right) + b_{12}
U_1 U_2 - \tan \chi \left( B_{12} + B_{21} \right)
\label{U13}
\end{eqnarray}
with $U_1=B_{11},U_2=B_{22}$,  $B_{nm}$ from (\ref{T5}) and $b_{12}$ from (\ref{U6}). In the limit $\chi\to 0$, we recover the bound state
of twisted kinks, see (\ref{U5}). The chiral twist of the solution is time independent and can be inferred from the prefactors of the
$U_1U_2$ terms. It does not depend on $\chi$ and therefore coincides with the sum of the individual twists, like for the bound state.
Consider the oscillating terms in ${\cal N}$ and ${\cal D}$ first, i.e., those multiplied by $\tan \chi$. Using ordinary coordinates to 
exhibit their space and time dependence, the factors multiplying $\tan \chi$ can be cast into the form
\begin{eqnarray}
\left(\frac{\zeta_2}{\zeta_1^*} B_{12} + \frac{\zeta_1}{\zeta_2^*} B_{21}\right) & = & e^{-i(\varphi_1+\varphi_2)} (B_{12}+B_{21}),
\nonumber \\
(B_{12} + B_{21}) & = & - \frac{2e^{Kx}}{\sqrt{K^2+\Omega^2}} \sin \left( \Omega t  + \arctan \frac{K}{\Omega} \right),
\label{U14}
\end{eqnarray}
where we have introduced a wave number $K$ and frequency $\Omega$ generalizing the corresponding quantities from the (real) DHN
breather,
\begin{equation}
K = \sin \varphi_1 + \sin \varphi_2, \qquad \Omega = \cos \varphi_1 - \cos \varphi_2.
\label{U15}
\end{equation}
The period of the twisted breather is $T= 2\pi/\Omega$. The time independent parts of ${\cal N}$ and ${\cal D}$ can be evaluated with the 
help of 
\begin{equation}
U_1= \frac{e^{2x \sin \varphi_1}}{2 \sin \varphi_1}, \qquad U_2=\frac{e^{2x \sin \varphi_2}}{2 \sin \varphi_2},\qquad b_{12}=
\frac{1-\cos(\varphi_1-\varphi_2)}{1-\cos(\varphi_1+\varphi_2)}.
\label{U16}
\end{equation}
Let us now turn to the issue of self-consistency. Following the steps leading to (\ref{T30}), we
write $\omega$ in its Cholesky factorized form

\begin{eqnarray}
\omega =
\begin{pmatrix}
{ \sec \chi & \tan \chi \cr
\tan \chi & \sec \chi
}
\end{pmatrix}=L L^\dagger\quad, \quad 
L=
\begin{pmatrix}
{ \sqrt{\sec \chi} & 0 \cr
\sqrt{\cos \chi} \tan \chi & \sqrt{\cos \chi}
}
\end{pmatrix}
\label{eq:breather-omega1}
\end{eqnarray}
Then
\begin{eqnarray}
2 \left(L^\dagger M^{\dagger}\frac{1}{L^\dagger}+ \frac{1}{L} M L \right)=2
\begin{pmatrix}
{ 2\varphi_1 & -( \varphi_1 - \varphi_2) \tan \chi \cr 
-( \varphi_1 - \varphi_2) \tan \chi & 2 \varphi_2}
\end{pmatrix}
\end{eqnarray}
Using (\ref{T30}), the eigenvalues of this matrix give the two filling fractions as:
\begin{eqnarray}
\nu_\pm =\frac{\varphi_1+\varphi_2}{2\pi}\pm  \frac{\varphi_1-\varphi_2}{2\pi} \sec \chi .
\label{eq:n2-filling2}
\end{eqnarray}
The condition that $\nu_{\pm} \in [0,1]$ restricts the allowed range of $\chi$ for given twist angles $\varphi_1,\varphi_2$.

We illustrate these various examples in a few cases, using $(S,P)$ plots. In Fig.~\ref{FIG1}, a 2-kink bound state at rest 
(parameters: $\varphi_1=1.0,\, \varphi_2=0.8,\, \omega_{11}=3,\, \omega_{22}=1/\omega_{11}$) is shown. If one increases the distance
between the kinks by increasing $\omega_{11}$, one reaches eventually two static, non-interacting kinks which would show up as an open 
polygon made out of two of the straight line segments shown in Fig.~\ref{FIG0}. The breather with the same parameters as the 
2-kink and $\chi=1.1$ is illustrated in Fig.~\ref{FIG2}, where the different curves correspond to equidistant time steps. Fig.~\ref{FIG3} shows 
the scattering of two twisted kinks with $\varphi_1=1.0, \varphi_2=0.8$. The initial 
and final states consist of two straight line segments ending on the chiral circle. During the collision process (illustrated again by a
sequence of equidistant time steps), the kinks interchange their order. Clearly, these static pictures can give only an incomplete view
of the time dependent examples. A complete graphical representation requires animated plots, as provided in the ancillary files to 
this paper for the same parameters, see the Appendix and the files {\tt animation$\_$kink$\_$plus$\_$kink} and {\tt animation$\_$2-breather}.

\subsection{Three pole solutions}
In the preceding section we have discussed the TDHF solutions built out of two kinks in great detail. 
With increasing number of kinks (or poles in the complex $\zeta$-plane), both the number of different physical configurations and
the complexity of these solutions increase rapidly. It is straightforward to generate these solutions with Computer Algebra (CA) using the
general formalism and to check the self-consistency by a numerical diagonalization of a finite matrix. We will show examples of 
such calculations at the end of this and the following sections. We start with a survey of the different cases with 3 poles.

\begin{figure}[htb]
\includegraphics[scale=0.5]{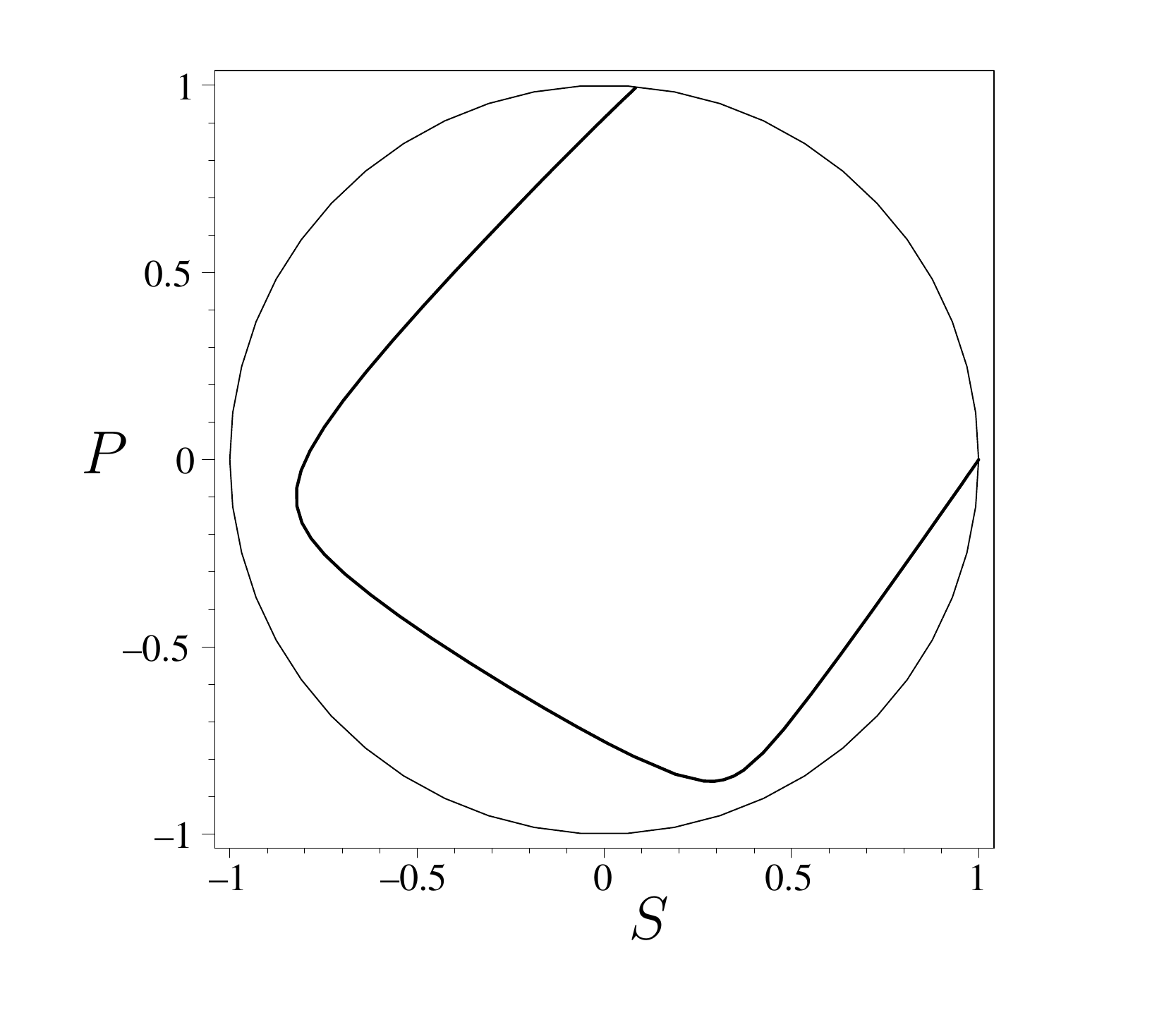}
\caption{($S,P$)-plot of a 3-kink, a bound state of 3 twisted kinks.}
\label{FIG4}
\end{figure}
\begin{figure}[htb]
\includegraphics[scale=0.5]{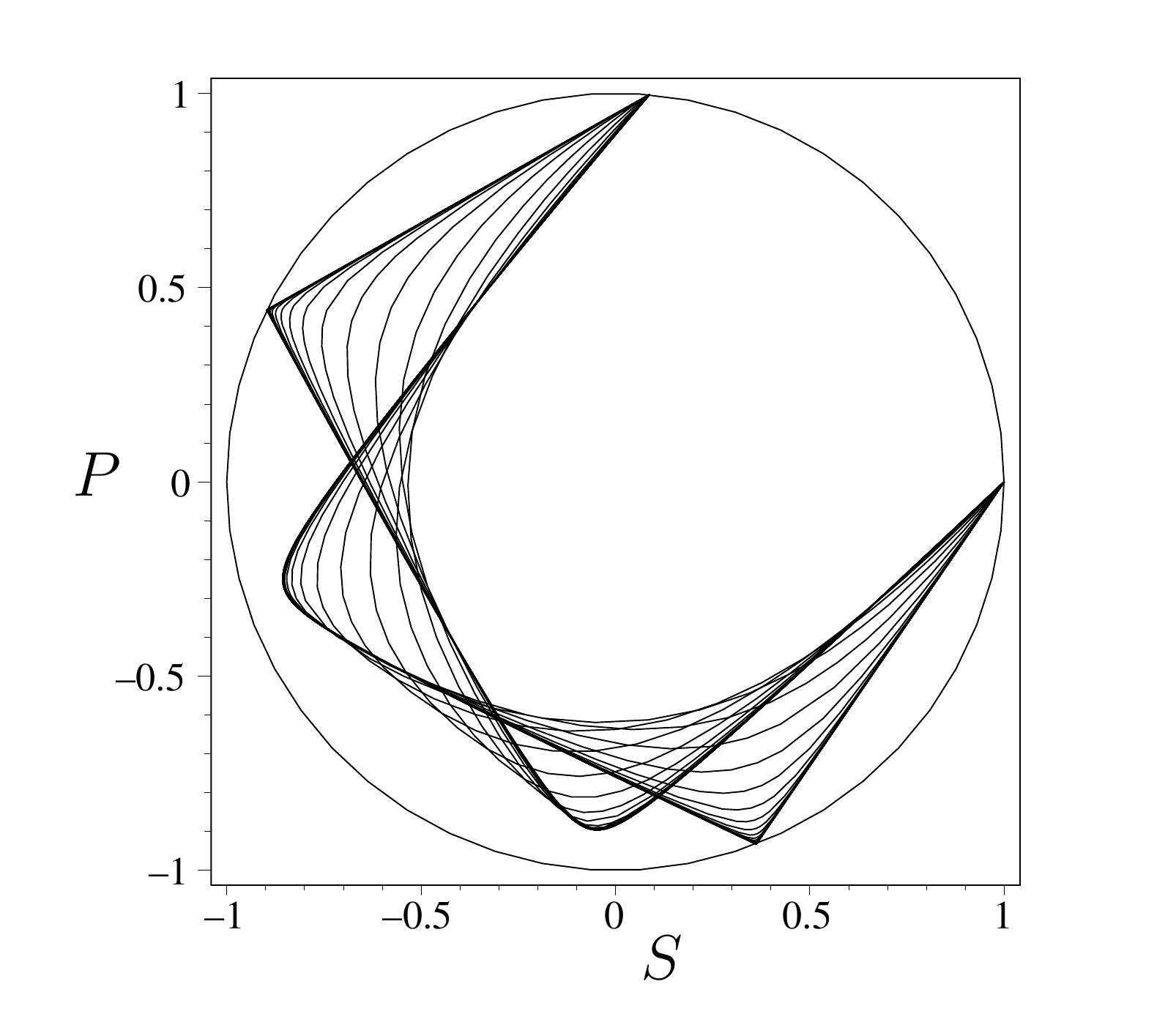}
\caption{($S,P$)-plot of the scattering of a 2-kink and a kink. The curves illustrate the time dependence, in equal time steps. Initial and final 
states can be identified by the fact that one inner point on a curve touches the chiral circle.}
\label{FIG5}
\end{figure}
\begin{figure}[htb]
\includegraphics[scale=0.5]{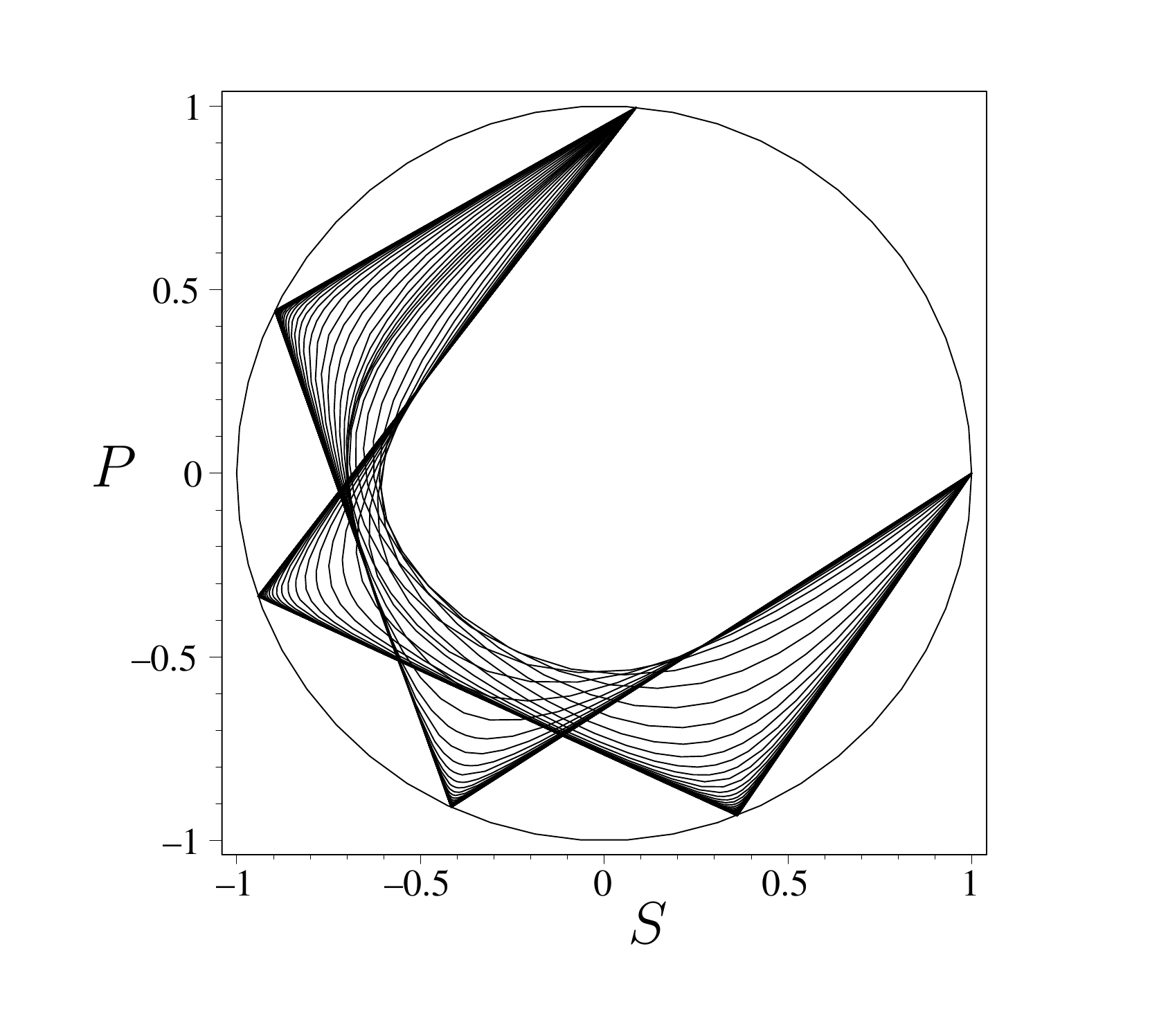}
\caption{$(S,P$)-plot of the scattering process of 3 single, twisted kinks. The curves show the time dependence, in equal time steps. Initial and final states correspond to 3-sided open 
polygons, all corners and endpoints lying on the chiral circle.}
\label{FIG6}
\end{figure}

The input to any TDHF calculation of the NJL$_2$ or GN models are a set of boost parameters $\eta_n$ and chiral twist angles $\varphi_n$ 
for the constituent kinks, together with the bound state mixing matrix $\omega$. These parameters are not entirely independent though. A
non-vanishing off diagonal
matrix element $\omega_{nm}$ implies that the physical bound states of kinks $n$ and $m$ get mixed. This is only physically meaningful
if these two kinks have the same boost parameter $\eta_n=\eta_m$, since otherwise the two kinks would be arbitrarily far apart at asymptotic
times and the mixing would violate cluster separability. The other restriction is that two kinks (not involved in breathers) with the same
$\eta_n$ parameter must have different $\varphi_n$'s, otherwise the number of kinks is reduced by 1.

With this in mind, the possibilities with 3 kinks are as follows. If $\eta_1,\eta_2,\eta_3$ are all different, we are dealing with the scattering
of 3 individual kinks. If one chooses in particular $\varphi_n=\pi/2$ for all 3 kinks, this reproduces known results for 3 CCGZ kinks of the
GN model
derived from the Sinh-Gordon solitons in \cite{Fitzner:2010nv}. If two of the kinks have the same velocity (say $\eta_1=\eta_2 \neq \eta_3$),
we are dealing with the scattering of a two-kink compound and a single kink and we must choose $\omega_{13}=\omega_{23}=0$.
The compound system can either be a bound state ($\omega_{12}=0$) or a breather ($\omega_{12} \neq 0$), as discussed in section 
\ref{sec:2poles}.
In the case of real potentials, this includes scattering and bound states of a DHN breather or baryon ($\varphi_2=\pi-\varphi_1$) and a 
CCGZ kink
($\varphi_3=\pi/2$). The bound state case has been discussed independently in the condensed matter \cite{Okuno:1983} and particle physics
\cite{Feinberg:2003} literature.
Finally, if all 3 kinks have the same velocity, there are 3 possibilities for $\omega$: If $\omega$ is diagonal, we describe the 3-kink bound
state which fits into the framework of \cite{Takahashi:2012pk}.
If only one off-diagonal element $\omega_{nm}$ is different from zero,
this describes a bound state of a 2-kink breather (kinks $n,m$) and a single kink. If more than one off-diagonal elements $\omega_{nm}$ are
different from zero, this three kink compound state cannot be resolved into a 2-kink breather and a kink, but represents a more
complicated oscillation mode where all 3 kinks are involved in a non-trivial way. Of course, in all of these cases one has to check that
the self-consistency condition can be fulfilled with physical occupation fractions $\nu_n\in [0,1]$. Since this involves diagonalization
of a 3$\times$3 matrix, this has to be checked on a case-by-case basis.

In order to simplify the discussion in the next section, we introduce the following language: A bound state of $n$ twisted kinks will be
referred to as ``$n$-kink" (a ``1-kink" being simply a kink). An irreducible breather made out of $n$ kinks will be called  ``$n$-breather". 
If several clusters are scattering, this will be indicated by a + sign, e.g., kink + kink for the scattering of 2 kinks.
Then the one pole solution deals with the kink, the two pole solution with kink + kink,  2-kink, 2-breather, and the three pole
solution with kink + kink + kink, kink + 2-kink, kink + 2-breather, 3-kink, 3-breather.

Let us illustrate once again a few cases, using $(S,P)$ plots. In Fig.~\ref{FIG4}, a 3-kink bound state at rest (parameters: $\varphi_1=1.0,\,
\varphi_2=0.8, \, \varphi_3=0.6,\,  \omega_{11}=9,\, \omega_{22}=1,\, \omega_{33}=1/9$) is shown.  
Fig.~\ref{FIG5} represents the scattering of a 2-kink bound state and a single kink ($\eta_1=\eta_2=2, \eta_3=1/2$), and Fig.~\ref{FIG6} the 
scattering 
of three twisted kinks ($\eta_1=2,\, \eta_2=1,\, \eta_3=1/2$). Similar plots involving 2-breathers or 3-breathers are not really able to convey a
picture of the complicated time dependence. We refer the reader to the Appendix and the ancillary files, where full animations of all of these cases can 
be found ({\tt animation$\_$kink$\_$plus$\_$kink$\_$plus$\_$kink, animation$\_$2-breather$\_$kink$\_$boundstate, animation$\_$2-breather$\_$plus$\_$kink, 
animation$\_$3-breather}). 

\subsection{Four pole solutions}
TDHF solutions based on four kinks are of particular interest, since we reach the level of complexity needed
to describe baryon-baryon and breather-breather scattering in the GN model. These problems have already been solved recently by a
different method based on an ansatz for the TDHF potential \cite{Dunne:2011wu,Fitzner:2012kb}, at the expense of a substantial
technical effort. It is an important cross-check of the present simpler approach to reproduce these complicated results. 

From the preceding discussion, it is clear that the various four kink processes can be classified as follows: kink + kink + kink + kink, 
kink + kink + 2-kink,
kink + kink + 2-breather, 2-kink + 2-kink, 2-kink + 2-breather, 2-breather + 2-breather, kink + 3-kink, kink + 3-breather, 4-kink, 4-breather.
Out of these, we select the following processes which are of interest for the GN model:
\begin{enumerate}

\item {\em 2-kink + 2-kink}

By pairing the twist angles ($\varphi_1+\varphi_2=\varphi_3+\varphi_4=\pi$) and using a diagonal matrix $\omega$, this particular
process can be
turned into scattering of two DHN baryons studied in \cite{Dunne:2011wu}. We have checked with CA that the present closed 
expressions reproduce
exactly the results of Ref.~\cite{Dunne:2011wu}, provided one chooses the origin of the $x$ and $t$ axes appropriately. 
This calculation can now be generalized to the scattering of two twisted 2-kinks in a straightforward manner.

\begin{figure}[htb]
\includegraphics[scale=0.5]{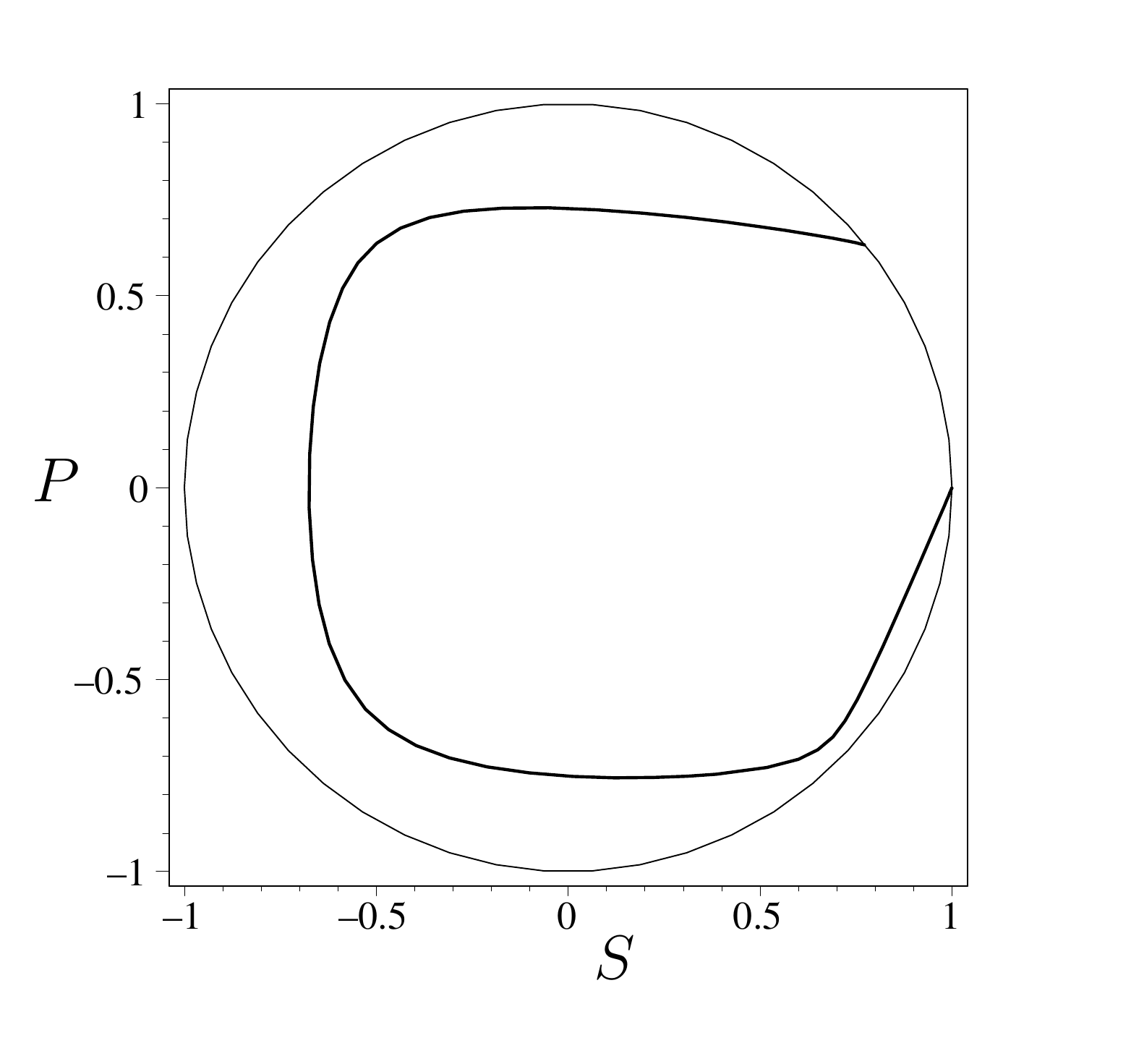}
\caption{($S,P$)-plot of a 4-kink, a bound state of four twisted kinks.}
\label{FIG7}
\end{figure}
\begin{figure}[htb]
\includegraphics[scale=0.5]{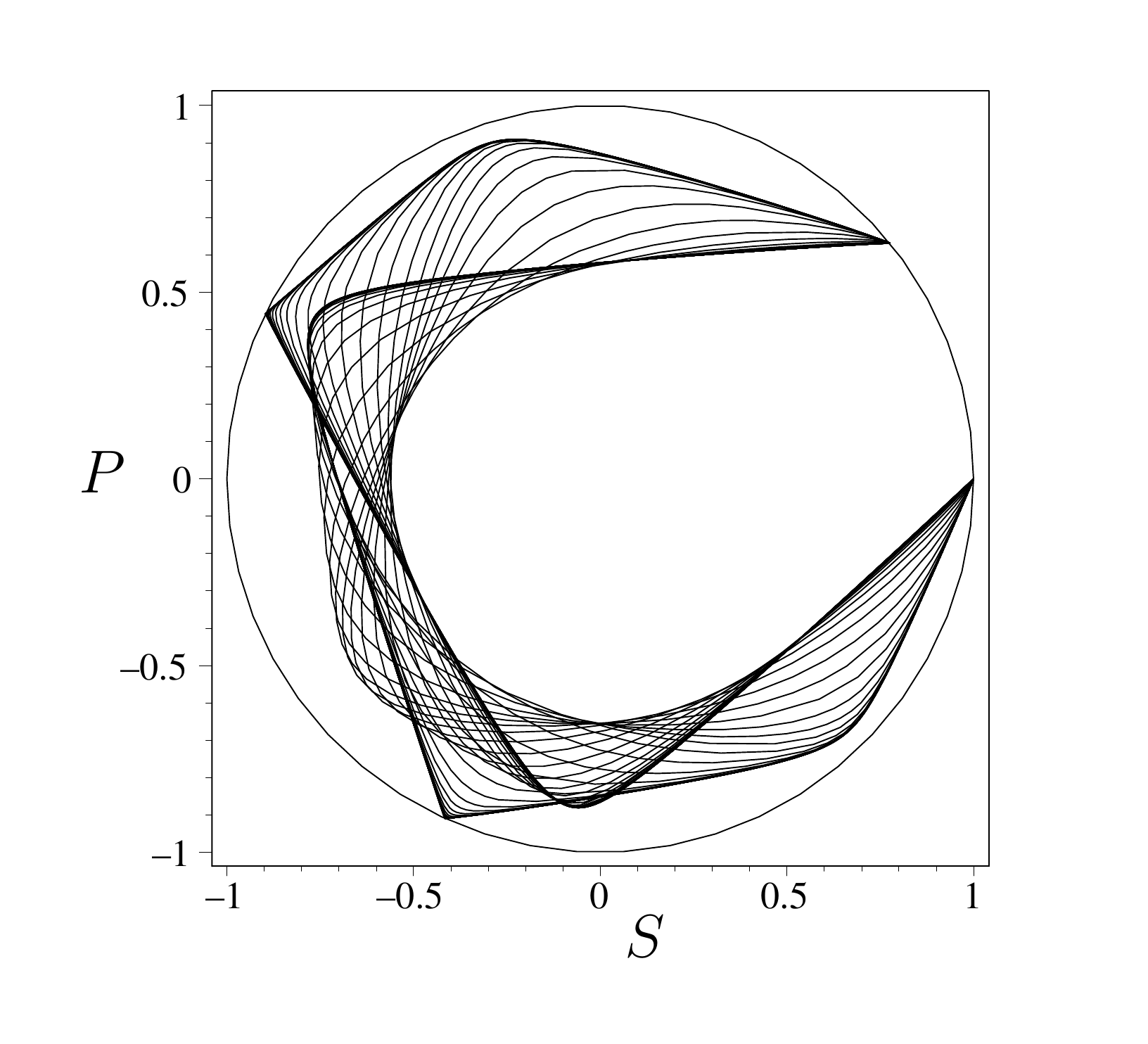}
\caption{($S,P$)-plot of the scattering process of two 2-kinks. The curves show the time dependence, in equal time steps. The initial and final states are the curves touching the chiral
circle with an inner point.} 
\label{FIG8}
\end{figure}

\item {\em 2-breather + 2-breather}

In order to get a real TDHF potential for breather-breather scattering, one has to pair the twist angles as in the baryon-baryon case 
and choose 
$\omega$ in the block diagonal form
\begin{equation}
\omega =
\begin{pmatrix}
{ \omega_{11} & \omega_{12} & 0 & 0 \cr
\omega_{12}^* & \omega_{11} & 0 & 0 \cr
0 & 0 & \omega_{33} & \omega_{34} \cr
0 & 0 & \omega_{34}^* & \omega_{33} 
}
\end{pmatrix}
\label{U17}
\end{equation}
Once again, we have checked with CA that the result agrees with the solution of breather-breather scattering in the GN model 
from \cite{Fitzner:2012kb}.
A comparison between the complicated formulas given in \cite{Fitzner:2012kb} and the present work shows how efficient it is to take 
the detour via the 
NJL$_2$ model, where one can take full advantage of factorization and integrability properties of the model. Once again, the present 
approach allows 
us to repeat the calculation with twisted breathers in the NJL$_2$ model with modest effort, solving an even more complicated problem 
analytically.

\item {\em 4-breather}

An irreducible four-kink breather of the NJL$_2$ model has many free parameters due to the appearance of a general, hermitean 
4$\times$4 matrix $\omega$. We do not study all of these complex oscillation modes here, but ask the question: How many parameters
survive if we specialize the 4-breather to real $\Delta$, i.e., a solution of the GN model? This is of some interest, since the 4-breather
is the simplest TDHF solution of the GN model which cannot be reduced to the known basic building blocks of kink, baryon and 2-breather.
(There is no real 3-breather, since the chiral twists have to be paired). We have computed the TDHF potential $\Delta$ for the 4-breather
at rest with CA, using $\varphi_1+\varphi_2=\varphi_3+\varphi_4=\pi$ and keeping $\omega$ general at first.
We then demand that $\Delta$ is real. This puts a number of constraints on the matrix elements $\omega_{nm}$. 
The most general solution can be parameterized as follows ($a$ and $e$ are real),
\begin{equation}
\omega = 
\begin{pmatrix}
{ a & b & c & d \cr b^*  & a & d^* & c^* \cr c^* & d & e & f  \cr d^* & c & f^*  & e 
}
\end{pmatrix}
\label{A58}
\end{equation}
This leaves a lot of room for new kinds of solutions of the GN model, parameterized by the 2 complex parameters $c,d$
characteristic for an irreducible 4-breather.

\end{enumerate}

In Fig.~\ref{FIG7}, a 4-kink bound state at rest (parameters: $\varphi_1=1.0,\,
\varphi_2=0.8,\, \varphi_3=0.6,\, \varphi_4=0.4,\,  \omega_{11}=81,\, \omega_{22}=9,\, \omega_{33}=1,\, \omega_{44}=1/9$) is illustrated.  
Fig.~\ref{FIG8} shows the scattering of a 2-kink on a 2-kink.  For animations of the complete time dependence and processes involving
breathers, see the Appendix and the ancillary files ({\tt animation$\_$2-kink$\_$plus$\_$2-kink, animation$\_$2-breather$\_$plus$\_$2-breather, animation$\_$4-breather}), where 
also an example of an irreducible 4-breather of the GN model with real $\Delta=S$ ({\tt animation$\_$real$\_$4-breather}) can be found.

\section{Summary and conclusions}
Within one year after the inception of the GN$_2$ model, DHN found a time dependent multi-fermion solution, the breather 
\cite{Dashen:1974ci}. They 
also realized that it is related to the kink-antikink scattering problem by analytic continuation. Somewhat surprisingly, no
further progress was made on time-dependent solutions of either the GN$_2$ or the NJL$_2$ model between 1975 and 2010, to the 
best of our knowledge. In the present work and in Refs.~\cite{Dunne:2013xta,Dunne:2013tr}, we have presented what we
believe to be the full solution of the TDHF problem for both  the GN$_2$ and NJL$_2$ models. Let us briefly summarize how this has 
been achieved.

In a first round of investigations starting in 2010, the interaction of a small number of scatterers was studied in great detail by
means of an ansatz method. The scatterers involved were kinks \cite{Klotzek:2010gp}, baryons \cite{Dunne:2011wu}  and
breathers \cite{Fitzner:2012kb}, all belonging to the GN$_2$ model. The ansatz consisted of multiplying the scalar potentials and spinors
for the individual scatterers and then varying the coefficients of some ($x,t$)-dependent exponentials, until the Dirac equation
was satisfied. This could be done at the expense of considerable use of computational algebra, and led to the exact 
solutions of the problems considered.
In the course of these works, many simplifying features emerged which enabled the authors to extrapolate the results to more
complicated scattering processes involving $N$ scatterers \cite{Fitzner:2012gg}. Since a general proof was lacking, these results
could only be checked analytically for few body problems, up to $N=6$. In the simplest special case, that of multi-kink 
scattering, the problem proved to be
fully solvable  for all $N$, by mapping it onto the known soliton solutions of the Sinh-Gordon equation \cite{Fitzner:2010nv}. 

Several developments have helped us to solve the problem in full generality in the meantime. Thus for instance, we realized that it is
advantageous to solve the NJL$_2$ model first, and then get the GN$_2$ solutions in a second step by specializing to real TDHF potentials. 
This strategy had been overlooked for a long time, and is indeed unexpected: the NJL$_2$ model has a more complicated Lagrangian 
than the GN$_2$ model. Moreover, its continuous chiral symmetry forbids states with localized fermion density, whereas one is just interested
in such ``baryonic" states in the GN$_2$ model. The reason why the NJL$_2$ model is easier to solve lies in the fact that twisted kinks
are the basic constituents of all TDHF solutions, and they appear in free form only in the NJL$_2$ model. Nevertheless, they are
also hidden constituents of GN$_2$ baryons and breathers, as we have shown here. As for the question of fermion 
density, we have shown that 
the same construction of the TDHF potential can be used for both models, but the self-consistency condition is different, leading to different
assignments of fermion number, but with the same condensate.

A two step procedure for solving the TDHF problem has proven most economic. In a first step, we have constructed a general
family of transparent scalar-pseudoscalar Dirac potentials \cite{Dunne:2013tr}, generalizing the method used for the stationary
Schr\"odinger equation by Kay and Moses long ago \cite{kaymoses}. This yields closed form expressions for processes involving $N$ 
twisted kinks.
Depending on the parameters, they describe kinks, bound states, breathers  and scattering processes among all of these
entities. In a second step reported in the present paper, we employ these transparent potentials in a TDHF calculation and
prove their self-consistency. While the method is completely general, it requires diagonalization of an  $N\times N$ matrix. 
Thus, for more than $N=2$ it is difficult to write general analytic expressions, so the  occupation fractions of the bound states are  
best determined numerically.
We have presented examples with up to 4 kinks, displaying a rich spectrum of scenarios,
in particular as far as breathers are concerned. If one specializes these examples to real potential, either by choosing the twist angle $\pi$
or by pairing two twisted kinks to total twist 0, one recovers all the preceding results from the GN$_2$ model. In contrast to the earlier
works, we now have the general proof of the Dirac equation and self-consistency condition, as well as compact closed expressions
in terms of determinants, valid for arbitrary numbers of  constituent  kinks. We have also learned that new kinds of breathers appear at each 
$N$, so that one cannot exhaust the dynamics of the GN$_2$ model via bound or scattering states of $N=2$ objects only. The basic
constituent common to all solutions is the twisted kink, which does not exist as a free entity in the GN model --- it is hidden.

Characteristic for integrable models is the fact that the transmission amplitude for a fermion on a compound object factorizes in the 
individual kink constituents. 
Nevertheless, there are nontrivial back-reaction effects which require fermion filling-fraction conditions for a
self-consistent TDHF solution.
We have shown that the factorized scattering translates into an additivity of the kink masses for all bound states
and breathers. It is also the key for finding the asymptotic behavior of the solitons after the  scattering has taken place. This
includes in general a deformation of the soliton shape, a time delay and (for breathers) a change in the phases of the oscillations.

Is this the end of the story?  Given the fact that all static HF solutions are known, the only loophole is  for the breathers. 
We have not yet completely ruled out that the ansatz we have used for finding transparent Dirac potentials misses some exotic breathers with an even 
more complicated structure. However, in view of the simplicity of the underlying Lagrangians, this  seems very unlikely.

\section{Appendix: Parameters used in the animations}

Here we collect the parameters used in the animations contained in the Ancillary/Supplementary Material to the present paper, which can be found at the link on this paper's arXiv page.

The fermion occupation fractions $\nu_n$ are not input, but the result of the self-consistency condition. For kinks, they can be 
computed as $\nu_n=\varphi_n/ \pi$, therefore they are not given below.
For solutions involving breathers, the fermion occupation numbers are derived from the eigenvalues in the consistency condition (\ref{T30}), as described at the end of
section \ref{sec:selfco}.
\begin{itemize}
\item animation$\_$kink$\_$plus$\_$kink
\begin{eqnarray}
& & \eta_1  =  2, \quad \eta_2 = 1/2
\nonumber\\
& & \varphi_1  =  1.0, \quad \varphi_2 = 0.8
\nonumber \\
& & \omega_{11} =   3, \quad \omega_{22} = 1/3 
\nonumber
\end{eqnarray}
\item animation$\_$2-breather
\begin{eqnarray}
& & \eta_1  =  \eta_2 = 1 
\nonumber \\
& & \varphi_1  =  0.6, \quad \varphi_2 = 1.2
\nonumber \\
& & \chi  =  1.1
\nonumber \\
& & \nu_1  =  0.076, \quad \nu_2 = 0.497 
\nonumber
\end{eqnarray}
\item animation$\_$kink$\_$plus$\_$kink$\_$plus$\_$kink
\begin{eqnarray}
& & \eta_1  =  2, \quad \eta_2 = 1, \quad \eta_3 = 1/2
\nonumber \\
& & \varphi_1  =  1.0, \quad \varphi_2 = 0.8, \quad \varphi_3 = 0.6
\nonumber \\
& & \omega_{11}  =   9, \quad \omega_{22} = 1, \quad \omega_{33} = 1/9
\nonumber
\end{eqnarray}
\item animation$\_$2-kink$\_$plus$\_$kink
\begin{eqnarray}
& & \eta_1  =  \eta_2 = 2, \quad \eta_3 = 1/2
\nonumber \\
& & \varphi_1  =  1.0, \quad  \varphi_2 = 0.8, \quad \varphi_3 = 0.6
\nonumber \\
& & \omega_{11}  =  9, \quad \omega_{22}=1, \quad \omega_{33}=1/9
\nonumber
\end{eqnarray}
\item animation$\_$2-breather$\_$kink$\_$boundstate
\begin{eqnarray}
& & \eta_1  =  \eta_2 = \eta_3 = 1
\nonumber \\
& & \varphi_1  =  1.0, \quad \varphi_2  =  0.8, \quad \varphi_3 = 0.6
\nonumber \\
& & \chi  =  1.4, \quad \omega_{33} = 1/81
\nonumber \\
& & \nu_1  =  0.474, \quad \nu_2 = 0.099
\nonumber
\end{eqnarray}
\item animation$\_$2-breather$\_$plus$\_$kink
\begin{eqnarray}
& & \eta_1  =  \eta_2 =  1.1, \quad \eta_3 = 1/\eta_1
\nonumber \\
& & \varphi_1  =  1.0, \quad \varphi_2  =  0.8, \quad \varphi_3 = 0.6
\nonumber \\
& & \chi  =  1.4, \quad \omega_{33} = 1/81
\nonumber \\
& & \nu_1  =  0.474, \quad \nu_2 = 0.099
\nonumber
\end{eqnarray}
\item animation$\_$3-breather
\begin{eqnarray}
& & \eta_1  =  \eta_2 =  \eta_3 = 1
\nonumber \\
& & \varphi_1  =  1.0, \quad \varphi_2  =  0.8, \quad \varphi_3 = 0.6
\nonumber \\
& & L  =  \left(\begin{array}{rrr} 2.5 & 0 & 0 \\ 2.4 & 0.5 & 0 \\ 1.65 & -0.7 & 1.0 \end{array} \right)
\nonumber \\
& & \nu_1  =  0.543, \quad \nu_2 = 0.007, \quad \nu_3 = 0.214
\nonumber
\end{eqnarray}
\item animation$\_$2-kink$\_$plus$\_$2-kink
\begin{eqnarray}
& & \eta_1  =  \eta_2 =  2, \quad \eta_3 = \eta_4 = 1/2
\nonumber \\
& & \varphi_1  = 1.0, \quad \varphi_2  =  0.8, \quad \varphi_3 = 0.6, \quad \varphi_4=0.4
\nonumber \\
& & \omega_{11}  =  81, \quad \omega_{22} = 9, \quad \omega_{33} = 1, \quad \omega_{44} = 1/9
\nonumber
\end{eqnarray}
\item animation$\_$2-breather$\_$plus$\_$2-breather
\begin{eqnarray}
& & \eta_1  =  \eta_2 =  1.1, \quad \eta_3 = \eta_4 = 1/\eta_1
\nonumber \\
& & \varphi_1  =  1.0, \quad \varphi_2  =  0.8, \quad \varphi_3 = 0.6, \quad \varphi_4 = 0.4
\nonumber \\
& & \chi_1  =  1.1, \quad \chi_2  = 1.2
\nonumber \\
& & \nu_1  =  0.357, \quad \nu_2 = 0.216, \quad \nu_3  =  0.247, \quad \nu_4 = 0.071 
\nonumber
\end{eqnarray}
\item animation$\_$4-breather
\begin{eqnarray}
& & \eta_1  =  \eta_2 =  \eta_3 = \eta_4 = 1
\nonumber \\
& & \varphi_1  =  1.0, \quad \varphi_2  =  0.8, \quad \varphi_3 = 0.6, \quad \varphi_4=0.4
\nonumber \\
& & L  =  \left(\begin{array}{rrrr} 1.48 & 0 & 0 & 0 \\ 1.32 & 0.67 & 0& 0  \\ 0 & 1.50 & 1.66 & 0 \\ 0 & 0 & 1.55 & 0.60  \end{array} \right)
\nonumber \\
& & \nu_1  =  0.473, \quad \nu_2 = 0.029, \quad \nu_3 = 0.172, \quad \nu_4 = 0.217
\nonumber
\end{eqnarray}
\item animation$\_$real$\_$4-breather
\begin{eqnarray}
& & \eta_1  =  \eta_2 =  \eta_3 = \eta_4 = 1
\nonumber \\ 
& & \varphi_1  =  1.0, \quad \varphi_2  = \pi - \varphi_1, \quad \varphi_3 = 0.6, \quad \varphi_4= \pi - \varphi_3
\nonumber \\
& & L  =  \left(\begin{array}{rrrr} 1.07 & 0 & 0 & 0 \\ 0.51 & 0.94 & 0 & 0 \\ 0.22 & 0.20 & 1.20 & 0 \\ 0.28 & 0.10 & 0.86 & 0.83  \end{array} \right)
\nonumber \\
& & \nu_3 - \nu_1  =  0.922, \quad \nu_4-\nu_2 = 0.410
\nonumber
\end{eqnarray}
\end{itemize}

\section*{Acknowledgement}

G.D. acknowledges support from DOE grants DE-FG02-92ER40716 and DE-FG02-13ER41989, and the ARC Centre of Excellence in
Particle Physics at the Terascale and CSSM, School of Chemistry and Physics, University of Adelaide, and M.T. thanks the DFG for 
financial support under grant TH 842/1-1.
We thank P. Dunne for assistance with the animations, which are available in the Ancillary/Supplementary Material.

\end{document}